\documentclass[pdflatex,sn-mathphys-num]{sn-jnl}

\usepackage{graphicx}%
\usepackage{multirow}%
\usepackage{amsmath,amssymb,amsfonts}%
\usepackage{amsthm}%
\usepackage{mathrsfs}%
\usepackage[title]{appendix}%
\usepackage{xcolor}%
\usepackage{textcomp}%
\usepackage{manyfoot}%
\usepackage{booktabs}%
\usepackage{algorithm}%
\usepackage{algorithmicx}%
\usepackage{algpseudocode}%
\usepackage{listings}%

\theoremstyle{thmstyleone}%
\newtheorem{theorem}{Theorem}
\theoremstyle{thmstyletwo}%
\newtheorem{remark}{Remark}%

\theoremstyle{thmstylethree}%

\begin{document}

\title[Article Title]{The Study of the Canonical forms of Killing tensor in vacuum with $\Lambda$}


\author*[1]{\fnm{D.} \sur{Kokkinos}}\email{kokkinos@physics.uoc.gr}

\author[2]{\fnm{T.} \sur{Papakostas}}\email{taxiarchis@hmu.gr}


\affil*[1]{\orgdiv{Department of Information and Communication Systems Engineering}, \orgname{University of the Aegean}, \orgaddress{\city{Karlovasi}, \state{Samos}, \country{Greece}}}

\affil[2]{\orgdiv{Department of Electrical and Computer Engineering}, \orgname{Hellenic Mediterranean University}, \orgaddress{ \city{Heraklion}, \state{Crete}, \country{Greece}}}



\abstract{ This paper is the initial part of a comprehensive study of spacetimes that admit the canonical forms of Killing tensor in General Relativity. The general scope of the study is to derive either new exact solutions of Einstein's equations that exhibit hidden symmetries or to identify the hidden symmetries in already known spacetimes that may emerge during the resolution process.
In this preliminary paper, we first introduce the canonical forms of Killing tensor, based on a geometrical approach to classify the canonical forms of symmetric 2-rank tensors, as postulated by R. V. Churchill. Subsequently, the derived integrability conditions of the canonical forms serve as additional equations transforming the under-determined system of equations, comprising of Einstein's Field Equations and the Bianchi Identities (in vacuum with $\Lambda$), into an over-determined one. Using a null rotation around the null tetrad frame we manage to simplify the system of equations to the point where the geometric characterization (Petrov Classification) of the extracted solutions can be performed and their null congruences can be characterized geometrically. Therein, we obtain multiple special algebraic solutions according to the Petrov classification (D, III, N, O) where some of them appeared to be new. The latter becomes possible since our analysis is embodied with the usage of the Newman-Penrose formalism of null tetrads.} 






\maketitle


\section{Introduction}\label{sec1}
The objective of this paper is to find spacetimes with hidden symmetries, which are exact solutions of Einstein's equations in vacuum with cosmological constant $\Lambda$, using the existence of an abstract 2-rank Killing tensor as the only premise. It should be noted that no additional assumptions are made regarding specific Petrov types, invertibility, separability, or other conditions. For reasons of clarity, we begin with a review in which our motives, goals and outcomes will be further explained, providing the reader with a broader understanding of the scope of this work. 

The interpretation and the analysis of exact solutions in the context of General Relativity constitute a whole regime of research. In this scientific branch exact solutions represent the hidden trophy within the non-linear nature of the equations of gravity. These solutions can be obtained by imposing restrictions on the under-determined system of equations consisting of Einstein's Field Equations (EFE) and Bianchi Identities (BI)\footnote{Since this work was operated in the framework of Newman-Penrose formalism, the EFE corresponds to the Newman-Penrose Equations (NPE).}. Such restrictions may be related to the algebraic structure of the Riemann tensor, to symmetry conditions on the metric, to the field equations of the stress energy-momentum tensor and other related considerations.   

The appropriate way to obtain exact solutions based on the bespoken assumption is to solve a system of equations regarding the EFE, BI and the Integrability Conditions (IC) of each canonical form of Killing tensor. According to this, we begin by introducing the canonical forms of the Killing tensor and deriving the integrability conditions for each of them. Hence, the resolution process initiates by constructing an over-determined-solvable system of equations consisting of EFE, BI, and the IC of each Killing form. The latter description makes clear that the canonical forms is the cornerstone of this problem.

The derivation of the canonical forms of a symmetric matrix will be proved a quite challenging task if someone attempts to do it algebraically. Churchill in 1932 \cite{churchill1932canonical} presented the canonical forms of a symmetric 2-rank tensor in pseudo-Euclidean spacetime based on a geometrical approach. This geometrical approach is based on 1) a theorem of Rainich regarding the existence of invariable planes within spacetime \cite{rainich1925electrodynamics} and 2) the existence of null vectors within these planes. We managed to acquire the canonical forms of Killing tensor in the framework of Newman-Penrose formalism \cite{Newman1962} following the exact same procedure as that operated by Churchill, considering an opposite signature though (Chapter \ref{section 5}). 

The next step in the resolution process is to determine the system of equations (EFE, BI, IC) and to operate a null rotation around the null tetrad frame wherein the Killing form must remains invariant. This transformation was used to constrain the arbitrariness of the null tetrad frame while simultaneously simplifying the system of equations.

In this preliminary work, the resolution procedure stopped at the point where the classification of the gravitational field according to Petrov \cite{petrov2000classification} and the characterization of the null congruences take place resulting in to six theorems. We managed to extract all the non-conformally flat solutions in vacuum with $\Lambda$ using as an assumption the existence of canonical forms of Killing tensor. As a matter of fact, in Chapters \ref{section 8}, \ref{section 9} the type D solutions with a shear-free and non-geodesic null congruence and with a shearing and geodesic null congruence that described by Theorems \ref{theorem 4}, \ref{theorem 5} accordingly, appear to be new as we claim in Chapter \ref{Section 10}. We intuit that they could be possible reductions (in vacuum with $\Lambda$) of the undiscovered solutions with Principal Null Directions of Einstein-Maxwell tensor aligned (or non-aligned) with the Principal Null Directions of Weyl tensor \cite{van2017algebraically}. Apparently, this work paves the way to find their spacetime metrics in full detail by solving the remaining equations, but first determining a coordinate system using the Frobenius theorem of integrability.

It should be noted that the main idea that motivated us to deal with the canonical forms of Killing tensor was \textit{that we might find new interesting spacetimes in vacuum if we deal with more general forms of a Killing tensor with more than two distinct eigenvalues} as postulated by Hauser and Malhiot in \cite{Hauser1976}. During the last decades the only works in the literature that utilize a Killing tensor to explore new spacetimes including Hauser and Malhiot's work on electro-vacuum \cite{hauser1978forms} and the work of one of us on interior solutions with perfect fluid \cite{papakostas1998generalization}. Hauser-Malhiot managed to found one of the most general family of stationary axially symmetric electro-vacuum spacetimes that found independently by Carter \cite{carter1968hamilton}. But both of these works serve as the only \textit{paradigms} for our research, since these are the only works both originate from the assumption of existence of a Killing tensor with two double eigenvalues only. 
It is worth mentioning that the Killing tensor form with two double eigenvalues is a special case of the canonical forms of a Killing tensor, and its study has yielded general and new families of exact solutions in electro-vacuum.      

\begin{equation}\tag{Paradigm} K_{\mu \nu} = \begin{pmatrix}
0 & \lambda_1   & 0 & 0\\
 \lambda_1 & 0 & 0 & 0\\
0 & 0&  0   &\lambda_2 \\
0 & 0 & \lambda_2 & 0  
\end{pmatrix}\end{equation}

It is both interesting and important to explore Einstein's spacetimes that admit more general forms of Killing tensors beyond the case of two double eigenvalues as initial premise. It is also intriguing to investigate whether the canonical forms can lead to new solutions or generalizations of the ones already known. Additionally, in the future we aim to investigate whether more general Petrov type solutions (e.g. Petrov type I) admit the canonical forms of the Killing tensor, as opposed to the Killing form of the \textit{Paradigm}. 

We are intrigued to study spacetimes with hidden symmetries derived from Killing tensors, because there are indications related to the physical nature of the trajectories within these spacetimes. It is crucial to consider that the observed trajectories are closed. A relevant conjecture suggests that in a spacetime admitting a non-trivial Killing tensor\footnote{The \textit{trivial} Killing tensor is the metric tensor $g_{\mu \nu}$ where its existence indicates the conservation of the rest mass of a moving particle in Hamiltonian systems.}, closed trajectories are present\cite{burns2021open}. 

Furthermore, as Eisenhart \cite{eisenhart1934separable} and Kalnins-Miller \cite{kalnins1980killing}, \cite{kalnins1981killing}, \cite{kalnins1983conformal} showed, the geodesic separation is correlated with the existence of Killing vectors and Killing tensors of valence two \cite{benenti2016separability} in some cases, leads to the separation of Hamilton-Jacobi equation providing us with integrable trajectories. Regarding this, the assumption of existence of Killing tensor could serve as a promising starting point in the pursuit of ``realistic" spacetimes endowed by integrable trajectories.

For instance, a well known example of a hidden symmetry is the conservation of Laplace-Runge-Lenz vector in the Kepler-Coulomb problem along geodesics as a constant of motion which is connected with the existence of Killing-Yano Tensor \cite{Papakostas2001}, a generalization of Killing Tensor \cite{cariglia2014hidden},\cite{taxiarchis1985space}. Another well known example of this kind of symmetry is the Carter's constant which is associated with probe inclination \cite{Carter1968b} and it is considered the fourth constant of motion in Kerr-like geometry.

Consequently, entangling the Killing tensor in a physical problem we achieve a twofold result: on one hand, by capitalizing on the Killing tensor, we incorporate additional equations (IC) into the previously under-determined system (EFE, BI), thereby transforming it into an over-determined one; on the other hand, the resulting spacetimes are endowed with hidden symmetries.

It's also worth noting that the use of the standard metric formalism is not practical for our purpose, as it cannot accommodate the symmetries associated with a Killing tensor. In this context, the most suitable formalism is the complex vectorial formalism of spin coefficients \cite{cahen1967complex} or the Newman-Penrose formalism \cite{newman1962approach}. The first reason is that during the resolution process, the classification of the gravitational field of a spacetime according to Petrov occurs in the early stages of the process. This allows us to deduce theorems of symmetries for each Petrov type, determining an appropriate coordinate system using the Frobenius theorem.

Secondly, capitalizing on the Killing equations we obtain simplifications between the spin coefficients and additional relations to Newman-Penrose field equations. The latter plays a pivotal role in solving the field equations without the need for specific coordinate system. This is possible since the Newman-Penrose field equations are first-order differential equations of spin coefficients. In fact, the complex vectorial formalism we employ provides insights into the essential characteristics of null congruences (Shear, Divergence, Geodesic), which are related to singularity theorems.


We have attempted to establish a coherent structure for this work. In Chapter \ref{section 3} we will exhibit the main points of the appointed formalism which will be revisited throughout the paper. Next, in Chapter \ref{section 4} we imply a rotation around the null tetrad frame with $l^\mu$ fixed and we obtain the general key relations that can be applied to any canonical form except $K^0$.  In Chapter \ref{section 5} the definition of Killing tensor is given and we introduce the canonical forms of Killing tensor in the framework of Newman-Penrose formalism based on Churchill's work . After that, in Chapters \ref{section 6}, \ref{section 7} and \ref{section 8} the resolution process takes place for $K^0$, $K^1$ and $K^{2,3}$, therein we present the Petrov types of the obtained solutions and we determine the characteristics of their null congruences. The Chapter \ref{section 9} before \textit{Discussion and Conclusions} contains a reduced Killing form with $\lambda_0=0$ which is a subcase of $K^1,K^2,K^3$ Killing forms. In the final chapter of this work (Chapter \ref{Section 10}), we briefly discuss, among other things, the role of entangled transformations in the resolution process (Lorentz transformations and null rotations) with the aim of obtaining exact solutions for more general Petrov types.

 


\section{Notation of the Newman-Penrose Formalism}\label{section 3}

The Newman-Penrose Formalism is a widely known formalism that was presented by Newman and Penrose \cite{newman1962approach} and was analyzed geometrically by Cahen, Debever and Defrise \cite{cahen1967complex}, \cite{debeverriemann}. Initially, the formalism was found in order to describe the gravitational radiation in General Relativity but it was proved to have much more usefulness.

The main concept of the formalism could be briefly described as follows. \textbf{The need to interpret the gravitational radiation more conveniently forces us to associate the Riemann tensor with isotropic null tetrads (light-like vectors)}. The latter could happen in a 3-dimensional complex bivector space ($C_3$) spanned by self-dual 2-forms. 

\subsection{Coordinate system}

The metric can be put in the form

\begin{equation}ds^2 = 2(\boldsymbol{\theta}^1 \boldsymbol{\theta}^2 - \boldsymbol{\theta}^3 \boldsymbol{\theta}^4) \end{equation}
where the general metric $g_{\mu \nu}$ is the following and equal to its inverse $g^{\mu \nu}$.

\begin {equation}g_{\mu \nu} = l_\mu n_\nu + n_\mu l_\nu   - m_\mu \bar{m}_\nu - \bar{m}_\mu m_\nu  = \begin{pmatrix}
0 &1&0&0\\
1&0&0&0\\
0&0&0&-1\\
0&0&-1&0
\end{pmatrix}\end{equation}
The pseudo-orthonormal basis contains two real and two complex conjugate vectors

\begin{equation} \boldsymbol{\theta}^1 \equiv n_\mu dx^\mu \hspace{0.8cm} \boldsymbol{\theta}^2 \equiv l_\mu dx^\mu \hspace{0.8cm} \boldsymbol{\theta}^3 \equiv - \bar{m}_\mu dx^\mu \hspace{0.8cm} \boldsymbol{\theta}^4 \equiv - m_\mu dx^\mu  \end{equation}
and the orthogonality properties of the vector components are the following since the rest combinations give zero.

\begin{equation}\label{orthog prop} l_\mu n^\mu = 1 = - m_\mu \bar{m}^\mu \end{equation}
The directional derivatives (dual basis) of the formalism are given by

$$\boldsymbol{D}  =  l^{\mu} \partial_\mu  \hspace{0.8cm}\boldsymbol{\Delta} = n^{\mu} \partial_\mu \hspace{0.8cm}\boldsymbol{\delta}  = m^{\mu} \partial_\mu \hspace{0.8cm} \boldsymbol{\bar{ \delta}} = \bar{m}^{\mu} \partial_\mu $$
Using the Cartan's method we can calculate the connection 1-forms ${\boldsymbol{\Gamma}^\alpha}_\nu \equiv {\Gamma^\alpha}_{\mu \nu}  \boldsymbol{\theta}^\mu$. 

\begin{equation}  d\boldsymbol{\theta}^\alpha  = - {\boldsymbol{\Gamma}^\alpha}_\nu \wedge  \boldsymbol{\theta}^\nu   \end{equation}
which is explicitly written as follows

\vspace{0.1cm}

\footnotesize
\begin{equation}\label{dtheta1} d\theta^1 = (\gamma+\bar{\gamma}) \theta^1 \wedge \theta^2 +(\bar{\alpha}  +\beta - \bar{\pi}) \theta^1\wedge \theta^3 +(\alpha+\bar{\beta} - \pi)\theta^1 \wedge \theta^4 -\bar{\nu}\theta^2\wedge \theta^3 - \nu\theta^2\wedge \theta^4 -(\mu-\bar{\mu})\theta^3\wedge \theta^4   \end{equation}
\begin{equation}\label{dtheta2} d\theta^2 = (\epsilon+\bar{\epsilon}) \theta^1 \wedge \theta^2 +\kappa \theta^1\wedge \theta^3 +\bar{\kappa}\theta^1 \wedge \theta^4 -(\bar{\alpha}+\beta -\tau)\theta^2\wedge \theta^3 - (\alpha+\bar{\beta} -\bar{\tau})\theta^2\wedge \theta^4 -(\rho-\bar{\rho})\theta^3\wedge \theta^4   \end{equation}
\begin{equation}\label{dtheta3} d\theta^3 = -(\bar{\tau}+\pi) \theta^1 \wedge \theta^2 - (\bar{\rho} +\epsilon-\bar{\epsilon}) \theta^1\wedge \theta^3 -\bar{\sigma}  \theta^1 \wedge \theta^4 +(\mu -\gamma+\bar{\gamma})\theta^2\wedge \theta^3   +\lambda      \theta^2\wedge \theta^4 +(\alpha-\bar{\beta})\theta^3\wedge \theta^4   \end{equation}
\begin{equation}\label{dtheta4} d\theta^4 = -(\tau+\bar{\pi}) \theta^1 \wedge \theta^2 -\sigma  \theta^1 \wedge \theta^3- (\rho- \epsilon+\bar{\epsilon}) \theta^1\wedge \theta^4  +\bar{\lambda}      \theta^2\wedge \theta^3 + (\bar{\mu} +\gamma-\bar{\gamma})\theta^2\wedge \theta^4    -(\bar{\alpha}-\beta)\theta^3\wedge \theta^4   \end{equation}
\normalsize
the greek letters represent the 12 complex spin coefficients. In Newman-Penrose formalism the Christoffel symbols are represented by the spin coefficients or spin connections. 
\normalsize
The relations (\ref{dtheta1})-(\ref{dtheta4}) are obtained by the usage of the covariant derivatives of the null tetrads. 
\footnotesize

\begin{multline}\label{nderivative} n_{\mu;\alpha} = -(\epsilon + \bar{\epsilon})n_\alpha n_\mu -(\gamma+\bar{\gamma})l_\alpha n_\mu +(\alpha+\bar{\beta})m_\alpha n_\mu + (\bar{\alpha}+\beta)\bar{m}_\alpha n_\mu + \pi n_\alpha m_\mu\\
 + \nu l_\alpha m_\mu -\lambda m_\alpha m\mu  -\mu \bar{m}_\alpha m_\mu  +\bar{\pi}  n_\alpha \bar{m}_\mu +\bar{\nu} l_\alpha \bar{m}_\mu -\bar{\mu} m_\alpha \bar{m}_\mu -\bar{\lambda} \bar{m}_\alpha \bar{m}_\mu\end{multline}
\begin{multline}\label{lderivative} l_{\mu;\alpha} = (\epsilon + \bar{\epsilon})n_\alpha l_\mu +(\gamma+\bar{\gamma})l_\alpha l_\mu -(\alpha+\bar{\beta})m_\alpha l_\mu - (\bar{\alpha}+\beta)\bar{m}_\alpha l_\mu - \bar{\kappa}n_\alpha m_\mu \\
- \bar{\tau}l_\alpha m_\mu + \bar{\sigma} m_\alpha m\mu +\bar{\rho} \bar{m}_\alpha m_\mu  - \kappa n_\alpha \bar{m}_\mu -\tau l_\alpha \bar{m}_\mu +\rho m_\alpha \bar{m}_\mu + \sigma \bar{m}_\alpha \bar{m}_\mu\end{multline}
\begin{multline}\label{mderivative} m_{\mu;\alpha} = - \kappa n_\alpha n_\mu - \tau l_\alpha n_\mu +\rho m_\alpha n_\mu +\sigma \bar{m}_\alpha n_\mu +\bar{\pi} n_\alpha l_\mu +\bar{\nu} l_\alpha l_\mu - \bar{\mu} m_\alpha l_\mu \\
- \bar{\lambda} \bar{m}_\alpha l_\mu +(\epsilon -\bar{\epsilon})n_\alpha m_\mu +(\gamma - \bar{\gamma})l_\alpha m_\mu - (\alpha - \bar{\beta})m_\alpha m_\mu +(\bar{\alpha} - \beta)\bar{m}_\alpha m_\mu
\end{multline}

\normalsize






\subsection{Bivector Space}\label{3.1}
The antisymmetricity of the electromagnetic tensor $F_{\mu \nu}$ provides us with six independent components $ \mu \nu = 12, 13, 14, 23, 24,34$ which are considered as a bivector basis in the 6-dimensional linear space ($M_6$) and also as elements of the orthochronous\footnote{The orthochronous Lorentz Group does not have mirroring in the timelike direction.} Lorentz group $SO^+(1,3)$ since they generate a Lie algebra.  
The 6-dimensional Lie algebra SO(1,3) is isomorphic to a complex 3-dimensional Lie algebra SL(2,C). Hence, the irreducible representation of the complex Lie algebra of the Lorentz group is embodied by the self-dual bivector basis.




This basis is defined by the following relations

\footnotesize
\begin{equation}\boldsymbol{Z}^1 = \boldsymbol{\theta}^1 \wedge \boldsymbol{\theta}^3 = Z^1_{\alpha \beta} dx^\alpha \otimes dx^\beta \hspace{0.1cm}; \hspace{0.2cm}Z^1_{\alpha \beta} = - n_\alpha \bar{m}_\beta + n_\beta \bar{m}_\alpha \end{equation}
\begin{equation}\boldsymbol{Z}^2 = \boldsymbol{\theta}^1 \wedge \boldsymbol{\theta}^2 - \boldsymbol{\theta}^3 \wedge \boldsymbol{\theta}^4 = Z^2_{\alpha \beta} dx^\alpha \otimes dx^\beta \hspace{0.1cm}; \hspace{0.2cm} Z^2_{\alpha \beta} =  n_\alpha l_\beta - n_\beta  l_\alpha - \bar{m}_\alpha m_\beta + m_\alpha \bar{m}_\beta \end{equation}
\begin{equation}\boldsymbol{Z}^3 = \boldsymbol{\theta}^4 \wedge \boldsymbol{\theta}^2 = Z^3_{\alpha \beta} dx^\alpha \otimes dx^\beta \hspace{0.1cm};  \hspace{0.2cm}Z^3_{\alpha \beta} = - m_\alpha l_\beta + m_\beta l_\alpha \end{equation}
\normalsize

\vspace{0.3cm}
The composite of the metric \footnote{The latin letters a,b,..,k, take values 1,2,3.} in this base is 

\begin{equation} \gamma^{a b} = 4\left[{\delta^{a}}_{(1} {\delta^b}_{3)} -  {\delta^{a}}_{2} {\delta^b}_{2} \right] = \begin{pmatrix} 0&0&2\\
			    0&-4&0\\
			    2&0&0	\end{pmatrix}															    		   	
 \end{equation}

The complex connection 1-forms $\boldsymbol{\sigma}_{b}^a$ is produced by derivation of the basis $\boldsymbol{Z}^a$, i.e 
\begin{equation}d\boldsymbol{Z}^a = -\boldsymbol{\sigma}^{a}_b \wedge \boldsymbol{Z}^b \end{equation}
the vectorial connection 1-form $\boldsymbol{\sigma}_a$ is defined by 
\begin{equation} \boldsymbol{\sigma}^a_b = 8 \epsilon^{kac} \boldsymbol{\sigma}_k \gamma_{cb} 
\hspace{0.2cm} \Leftrightarrow \hspace{0.2cm} \boldsymbol{\sigma}_k = \frac{1}{8} \epsilon_{k a c} \gamma^{c b} \boldsymbol{\sigma}^{a}_{b} =\kappa_{k \mu} \boldsymbol{\theta}^\mu
\end{equation}
where $ \epsilon_{a b c} $ is the Levi-Civita tensor and the tetrad components $ \kappa_{k \mu} $ contain the 12 complex spin coefficients
\footnotesize
\begin{equation}\kappa_{k \mu} = \begin{bmatrix}  \kappa & \tau & \sigma & \rho \\ \epsilon & \gamma & \beta & \alpha \\ \pi & \nu & \mu & \lambda\\          \end{bmatrix}\end{equation} 
\normalsize
The complex curvature 2-forms $ \boldsymbol{\Sigma}^{b}_{d} $ are defined by 

\begin{equation}\boldsymbol{\Sigma}^{b}_{d} = d \boldsymbol{\sigma}_{d}^{b} + \boldsymbol{\sigma}^{b}_{g} \wedge \boldsymbol{\sigma}^{g}_{d} \hspace{0.2cm} \Leftrightarrow \hspace{0.2cm}   \boldsymbol{\Sigma}_{a} = \frac{1}{8} e_{a b g } \gamma^{g d } \boldsymbol{\Sigma}{^b}_d  \end{equation}
The corresponding expanding of $\boldsymbol{\Sigma}_a$ with respect to the basis of $ \left( \boldsymbol{Z}^a , \boldsymbol{\bar{Z}}^a \right)$ is given by

\begin{equation}\boldsymbol{\Sigma}_a =   (C_{a b} - \frac{1}{6} R \gamma_{ a b})\boldsymbol{Z}^b + E_{a \bar{b}} \boldsymbol{\bar{Z}}^b ,  \end{equation} 
where these quantities are related with the curvature components of the formalism
\footnotesize
\begin{equation} C_{ a b} = \begin{bmatrix}  \Psi_0 & \Psi_1 & \Psi_2 \\ \Psi_1 & \Psi_2 &\Psi_3 \\ \Psi_2 & \Psi_3 & \Psi_4\\          \end{bmatrix},  \hspace{1cm} E_{ a \bar{b}} = \begin{bmatrix}  \Phi_{00} & \Phi_{01} & \Phi_{02} \\ \Phi_{10} & \Phi_{11} &\Phi_{12} \\ \Phi_{20} & \Phi_{21} & \Phi_{22}\\          \end{bmatrix}\end{equation}  
\normalsize
In this formalism, the 10 Weyl's components are represented by the 5 complex scalar functions. 
\footnotesize
$$ \Psi_0 = C_{\kappa \lambda \mu \nu} l^\kappa m^\lambda l^\mu m^\nu $$
$$ \Psi_1 = C_{\kappa \lambda \mu \nu} l^\kappa n^\lambda l^\mu m^\nu $$
\begin{equation} \Psi_2 = \frac{1}{2}C_{\kappa \lambda \mu \nu} l^\kappa n^\lambda \left[ l^\mu n^\nu -    m^\mu \bar{m}^\nu  \right] \end{equation}
$$ \Psi_3 = C_{\kappa \lambda \mu \nu} n^\kappa l^\lambda n^\mu \bar{m}^\nu$$
$$ \Psi_4 = C_{\kappa \lambda \mu \nu} n^\kappa \bar{m}^\lambda n^\mu \bar{m}^\nu $$
\normalsize

The Ricci tensor components are represented by $E_{a \bar{b}}$ and they are divided in to real and complex components. 
All these quantities describe the main parts of the EFE. The EFE in this formalism are represented by the corresponding field equations which are known either as Newman-Penrose Equations (NPE) or as Ricci identities \cite{newman1962approach}. 

\footnotesize
\begin{equation}\tag{a} D \rho  - \bar{\delta} \kappa = {\rho}^2 +\sigma \bar\sigma + \rho ( \epsilon + \bar{\epsilon}) - \bar{\kappa} \tau - \kappa \left[2(\alpha +\bar{\beta}) + (\alpha - \bar{\beta}) - \pi\right] \end{equation}
\begin{equation}\tag{b} \delta \kappa - D\sigma = - (\rho +\bar\rho +3\epsilon -\bar\epsilon)\sigma +\kappa \left[ \tau - \bar{\pi} +2(\bar{\alpha} +\beta) - (\bar{\alpha} - \beta) \right] - \Psi _o  \end{equation}
\begin{equation}\tag{c} D\tau = \Delta \kappa + \rho(\tau + \bar{\pi}) + \sigma(\pi +\bar\tau)+ \tau(\epsilon - \bar{\epsilon}) -2\kappa \gamma - \kappa (\gamma + \bar{\gamma}) + \Psi_1  \end{equation}
\begin{equation}\tag{i}\label{i} D\nu - \Delta \pi = \mu(\pi + \bar{\tau})+ \lambda(\bar\pi+\tau) +\pi(\gamma - \bar{\gamma}) -2\nu \epsilon - \nu (\epsilon + \bar{\epsilon}) +\Psi_3  \end{equation}
\begin{equation}\tag{g} \bar{\delta} \pi - D\lambda = - \pi(\pi + \alpha - \bar{\beta}) - \bar\sigma\mu + \nu\bar{\kappa} +\lambda(3\epsilon -\bar\epsilon)  \end{equation}
\begin{equation}\tag{p}\label{p} \delta \tau -\Delta\sigma = \mu \sigma+\bar\lambda\rho+ \tau (\tau - \bar{\alpha} + \beta)-\sigma(3\gamma -\bar\gamma) -\bar{\nu} \kappa   \end{equation}
\begin{equation}\tag{h} D\mu - \delta\pi = \mu \bar{\rho} + \sigma\lambda+ \pi(\bar{\pi} - \bar{\alpha} +\beta) -\mu (\epsilon +\bar{\epsilon}) - \kappa \nu + \Psi_2 + 2\Lambda \end{equation}
\begin{equation}\tag{n} \delta\nu - \Delta \mu = \mu (\mu + \gamma + \bar{\gamma}) +\lambda\bar\lambda - \bar{\nu} \pi + \nu (\tau - 2(\bar{\alpha}+\beta) +(\bar{\alpha} -\beta) ) \end{equation}
\begin{equation}\tag{q} \Delta \rho - \bar{\delta} \tau = - (\bar{\mu} \rho +\sigma\lambda  )- \tau(\bar{\tau} + \alpha - \bar{\beta}) + \nu \kappa + \rho(\gamma + \bar{\gamma}) - \Psi_2 - 2\Lambda\end{equation}
\begin{equation}\tag{k}\label{k} \delta\rho -\bar\delta\sigma= \rho(\bar{\alpha} + \beta) - \sigma(3\alpha-\bar\beta) +\tau(\rho-\bar{\rho})+ \kappa(\mu-\bar{\mu}) - \Psi_1 \end{equation}
\begin{equation}\tag{m}\label{m} \bar{\delta} \mu -\delta\lambda = -\mu (\alpha +\bar{\beta}) -\pi (\mu- \bar{\mu}) - \nu (\rho-\bar{\rho}) -\lambda(\bar\alpha - 3\beta) + \Psi_3 \end{equation}
\begin{equation}\tag{d} D\alpha - \bar{\delta} \epsilon = \alpha(\rho + \bar{\epsilon} -2\epsilon) +\beta\bar\sigma- \bar{\beta}\epsilon -\kappa\lambda - \bar{\kappa}\gamma + \pi (\epsilon + \rho) \end{equation}
\begin{equation}\tag{e}\label{e} D \beta - \delta{\epsilon} = \sigma(\alpha+\pi) +\beta(\bar{\rho} - \bar{\epsilon}) -\kappa(\mu + \gamma) -\epsilon(\bar{\alpha} - \bar{\pi}) + \Psi_1\end{equation} 
\begin{equation}\tag{r} \Delta \alpha - \bar{\delta}\gamma = \nu(\epsilon+\rho)-\lambda(\tau+\beta) +\alpha( \bar{\gamma} - \bar{\mu}) +\gamma (\bar{\beta}- \bar{\tau}) - \Psi_3 \end{equation}
\begin{equation}\tag{o}\label{o} -\Delta \beta + \delta \gamma = \gamma(\tau -\bar{\alpha} - \beta) +\mu \tau -\sigma\nu - \epsilon \bar{\nu} - \beta( \gamma - \bar{\gamma} -\mu)\end{equation}
\begin{equation}\tag{l}\label{l} \delta \alpha - \bar{\delta}\beta = \mu \rho-\sigma\lambda +\alpha (\bar{\alpha} - \beta) - \beta(\alpha - \bar{\beta})+  \gamma(\rho - \bar{\rho}) +\epsilon (\mu-\bar{\mu})-\Psi_2 + \Lambda \end{equation}
\begin{equation}\tag{f}\label{f} D\gamma - \Delta \epsilon = \alpha(\tau + \bar{\pi}) + \beta( \bar{\tau} + \pi) - \gamma ( \epsilon+\bar{\epsilon}) - \epsilon (\gamma +\bar{\gamma})   + \Psi_2 - \Lambda - \kappa \nu +\tau \pi\end{equation}
\begin{equation}\tag{j} \bar{\delta}\nu -\Delta\lambda = \lambda(\mu +\bar\mu+3\gamma -\bar\gamma)- \nu\left[2(\alpha +\bar{\beta} ) + (\alpha - \bar{\beta})  + \pi - \bar{\tau}  \right] + \Psi_4\end{equation}
\normalsize

\vspace{0.2cm}
The Bianchi Identities (BI) are given by the following relations

\footnotesize
\begin{equation}\tag{I}\label{I} \bar{\delta} \Psi_0 - D \Psi_1 = (4\alpha - \pi )\Psi_0  - 2(2\rho +\epsilon)\Psi_1 +3\kappa \Psi_2 \end{equation}
\begin{equation}\tag{II}\label{II} \bar{\delta} \Psi_1 -D\Psi_2 = \lambda\Psi_0+ 2(\alpha - \pi)\Psi_1 -3\rho \Psi_2 +2\kappa \Psi_3 \end{equation}
\begin{equation}\tag{III}\label{III} \bar{\delta}\Psi_2 - D\Psi_3 = -3\pi \Psi_2 +2\lambda\Psi_1 +2(\epsilon - \rho) \Psi_3 +\kappa\Psi_4 \end{equation}
\begin{equation}\tag{IV}\label{IV} \bar{\delta}\Psi_3 - D\Psi_4 = -2(\alpha +2\pi)\Psi_3 +(4\epsilon - \rho)\Psi_4 + 3\lambda\Psi_2  \end{equation} 
\begin{equation}\tag{V}\label{V} \Delta \Psi_0 - \delta \Psi_1  = (4\gamma - \mu) \Psi_0 -2(2\tau +\beta)\Psi_1 +3\sigma\Psi_{2}  \end{equation}
\begin{equation}\tag{VI}\label{VI} \Delta \Psi_1 - \delta \Psi_2 = \nu \Psi_0 + 2(\gamma - \mu)\Psi_1  -3\tau\Psi_2 +2\sigma\Psi_3 \end{equation}
\begin{equation}\tag{VII}\label{VII} \Delta \Psi_2 - \delta \Psi_3 =\sigma\Psi_4+ 2\nu\Psi_1 -3\mu\Psi_2 +2(\beta-\tau)\Psi_3  \end{equation} 
\begin{equation}\tag{VIII}\label{VIII} \Delta \Psi_3 - \delta \Psi_4  = 3\nu \Psi_2 - 2(\gamma+2\mu) \Psi_3 +(4\beta-\tau)\Psi_4 \end{equation}

\normalsize

Also, the Lie bracket plays an important role to the theory, since the commutation relations emerged by its implication on the vectors $n^\mu, l^\mu, m^\mu,\bar{m}^\mu $. The proper definition reads as follows for an arbitrary vector basis.

\footnotesize
\begin{equation} [ \boldsymbol{e}_\mu , \boldsymbol{e}_\nu] = -2 {\Gamma^\sigma}_{[\mu \nu]} \boldsymbol{e}_\sigma  \end{equation}
\normalsize
The commutations relations (CR) of the theory are given by
\footnotesize
\begin{equation}\tag{CR1}\label{CR1} [n^\mu,l^\mu ] = [D,\Delta] =  (\gamma+\bar{\gamma})D + (\epsilon + \bar{\epsilon})\Delta - (\pi + \bar{\tau})\delta - (\bar{\pi} +\tau)\bar{\delta} \end{equation} 
\begin{equation}\tag{$CR2_+$}\label{CR2+} [(\delta+\bar{\delta}),D] = (\alpha + \bar{\alpha} + \beta + \bar{\beta} - \pi - \bar{\pi} )D + (\kappa+\bar{\kappa})\Delta - (\bar\sigma+\bar{\rho}+\epsilon -\bar{\epsilon})\delta - (\sigma+\rho - \epsilon + \bar{\epsilon})\bar{\delta}  \end{equation}
\begin{equation}\tag{$CR2_-$}\label{CR2-} [(\delta-\bar{\delta}),D] = (-\alpha + \bar{\alpha} + \beta - \bar{\beta} + \pi - \bar{\pi} )D + (\kappa-\bar{\kappa})\Delta - (\bar{\rho} -\bar\sigma+\epsilon -\bar{\epsilon})\delta + (\rho -\sigma - \epsilon + \bar{\epsilon})\bar{\delta}  \end{equation}
\begin{equation}\tag{$CR3_+$}\label{CR3+} [(\delta+\bar{\delta}),\Delta] = -(\nu +\bar{\nu})D + (\tau+\bar{\tau} - \alpha - \bar{\alpha} - \beta - \bar{\beta})\Delta +(\mu +\lambda -\gamma +\bar{\gamma})\delta +(\bar{\mu} +\bar\lambda +\gamma - \bar{\gamma})\bar{\delta} \end{equation}
\begin{equation}\tag{$CR3_-$}\label{CR3-}[(\delta-\bar{\delta}),\Delta] = -(\nu -\bar{\nu})D + (\tau-\bar{\tau} + \alpha - \bar{\alpha} - \beta + \bar{\beta})\Delta +(\mu-\lambda -\gamma +\bar{\gamma})\delta -(\bar{\mu}-\bar\lambda +\gamma - \bar{\gamma})\bar{\delta} \end{equation}
\begin{equation}\tag{$CR4$}\label{CR4} [\delta,\bar{\delta}] = -(\mu - \bar{\mu})D - (\rho-\bar{\rho}) \Delta + (\alpha - \bar{\beta})\delta - (\bar{\alpha} - \beta)\bar{\delta}\end{equation}
\normalsize
All the above sets of equations contribute to the NPE, the BI and the CR of the basis vectors. Despite the fact that we have to solve a considerably larger number of equations this formalism has great advantages. Gauge transformations of the tetrad can be used to simplify the field equations and we can easily extract invariant properties of the gravitational field (Petrov types) without using a coordinate basis \cite{stephani2009exact}. Also, it allows us to search for solutions with specific special features, such as the presence of one or two null directions that might be singled out by physical or geometric considerations.

\section{Null rotation}\label{section 4}

The IC along with the NPE end up to be a cumbersome system of equations, thus, there are various approaches to obtain simplifications for our problem. The most used method involves implementing transformations by applying a rotation within the null tetrad frame (Null rotation) or exploring different options among the spin coefficients determining the rotation parameters (Lorentz transformations). 

In this work we choose to take advantage of the conformal symmetry of a rotation around one of the real null vectors, namely we choose $l^\mu$ to be fixed, although we will exhibit both transformations in the relations below, wherein the exponentials correspond to Lorentz transformations and the parentheses correspond to null rotation \cite{stewart1993advanced}. We define the complex rotation parameters $t \equiv a +ib$ and $p\equiv c+id$.

$$\tilde{\theta}^1 = e^{-a}(\theta^1+p\bar{p}\theta^2 +\bar{p}\theta^3 +p\theta^4 ) = \tilde{n}_{\mu} dx^{\mu}$$
$$\tilde{\theta}^2 = e^a \theta^2 = \tilde{l}_{\mu} dx^{\mu}$$
$$ \tilde{\theta}^3 = e^{-ib}(\theta^3 + p\theta^2)= -\tilde{\bar{m}}_{\mu} dx^{\mu}$$
$$\tilde{\theta}^4 = e^{ib}(\theta^4 +\bar{p}\theta^2)= -\tilde{m}_{\mu} dx^{\mu}$$ 
Subsequently, we require that the Killing tensor remains invariant under rotation. For instance we present the following compact form where $q=0,\pm1$ for $K^1$ and $K^2,K^3$ accordingly because the null rotation is not applicable for $K^0$ form.

$$K^{1,2,3} = \lambda_0 (\tilde{\theta}^1 \otimes \tilde{\theta}^1 +q \tilde{\theta}^2 \otimes \tilde{\theta}^2 ) +\lambda_1(\tilde{\theta}^1 \otimes \tilde{\theta}^2+\tilde{\theta}^2 \otimes \tilde{\theta}^1) + \lambda_2(\tilde{\theta}^3 \otimes \tilde{\theta}^4+\tilde{\theta}^4 \otimes \tilde{\theta}^3) +\lambda_7(\tilde{\theta}^3 \otimes \tilde{\theta}^3+\tilde{\theta}^4 \otimes \tilde{\theta}^4)$$

It is easy for someone to prove that the only non-zero rotation parameter is $t$ in case where the diagonal elements of the tensor are absent. This is valid due to the existence of the cross terms $\tilde\theta^1 \otimes \tilde\theta^2$ and $\tilde{\theta}^3 \otimes \tilde{\theta}^4$. Hence, in the next chapters scoping to obtain simplifications for our spin coefficients we may annihilate either $\lambda_0$ or $\lambda_7$ since the absence of other elements like $\lambda_1$ or $\lambda_2$ do not contribute to our cause. Let's proceed with the implication of the rotation on the spin coefficients. As we observe, the only further transformations that can be applied are Lorentz transformations since $p$ is equal to zero due to the invariance of the Killing tensor. Along these lines, to achieve the most general simplification, we won't correlate the spin coefficients with each other while determining the non-zero rotation parameter $t$. Thus, the transformation we apply is solely the null rotation.

$$\tilde\tau = e^{ib}\tau \hspace{1cm} \tilde\pi =  e^{-ib}\pi$$
$$\tilde\rho = e^{a}\rho \hspace{1cm} \tilde\mu = e^{-a}\mu$$
$$\tilde\kappa = e^{2a+ib}\kappa \hspace{1cm} \tilde\nu =  e^{-(2a+ib)}\nu$$
$$\tilde\sigma = e^{a+2ib}\sigma \hspace{1cm} \tilde\lambda = e^{-(a+2ib)}\lambda$$
$$\tilde{\beta} = e^{ib}\left(\beta+\frac{\delta t}{2} \right)  \hspace{1cm} \tilde{\alpha} = e^{-ib}\left(\alpha +\frac{\bar{\delta}{t}}{2} \right)$$
$$\tilde{\epsilon} = e^a\left(\epsilon +\frac{D t}{2}\right) \hspace{1cm} \tilde{\gamma} = e^{-a}\left(\gamma+\frac{\Delta t}{2}\right)$$
There are two different kinds of simplifications that can be acquired by the capitalization of the annihilation of the tilded spin coefficients $\tilde\epsilon, \tilde\gamma, \tilde\alpha, \tilde\beta$. The simplest simplification emerges by the correlation of the spin coefficient with the derivative of the rotation parameter $t$. In case where $\lambda_0=0$ the rotation parameter becomes $t=a$, differently, in case where $\lambda_7=0$ the non-zero rotation parameter is $t=ib$. The latter has significant impact on the spin coefficients. When $t=a$ the spin coefficients $\epsilon, \gamma, \alpha, \beta$ depends on the directional derivatives of real quantities yielding the following relations

$$\epsilon-\bar\epsilon=0$$
$$\gamma-\bar\gamma = 0$$
$$\alpha-\bar\beta=0$$
On the other hand when $t=ib$ we get

$$\epsilon+\bar\epsilon=0$$
$$\gamma+\bar\gamma = 0$$
$$\alpha+\bar\beta=0$$
The second kind of simplification takes place when we substitute the four tilded spin relations into the CR and we compare the outcome with the NPE (\ref{f}),(\ref{l}),(\ref{e}),(\ref{o}) resulting in the \textit{key relations}. The key relations help us to unfold the branches of the solutions. We postulate the most general case of the obtained relations after the comparison with NPE.    

\begin{equation}\label{kri}\tag{i} \Psi_2 - \Lambda = \kappa \nu - \tau \pi \end{equation}
\begin{equation}\label{krii}\tag{ii}\Psi_1 = \kappa \mu - \sigma\pi\end{equation}
\begin{equation}\label{kriii}\tag{iii}\Psi_2 - \Lambda = \mu \rho - \sigma\lambda\end{equation}
\begin{equation}\label{kriv}\tag{iv}\mu \tau-\sigma\nu = 0\end{equation}
The null rotation provides us with these useful relations which connect the spin coefficients along with the Weyl components. This result depends mainly on the form of the Killing Tensor, since we demand the preservation of Killing tensor. Essentially, the lack either of $\lambda_0$ or $\lambda_7$ allows us to obtain simplifications such as the \textbf{key relations} \footnote{The annihilation of $\lambda_0$ serves as a possible reduction to this canonical form although the absence of $\lambda_0$ in $K^1$, $K^2$, $K^3$ makes them to coincide. This choice is exhibited in Chapter 9.}.

\begin{remark}\label{remark}
    The application of null rotation and Lorentz transformations are transformative processes that yield valuable relationships connecting not only the spin coefficients amongst themselves but also with the Weyl components through the commutation relations, once the tilded spin coefficients have been annihilated. The outcome critically hinges on the preservation of the structure of the Killing tensor and the form of the spin coefficients. Importantly, the absence of either $\lambda_0$ or $\lambda_7$ is the catalyst for these relationships. \textbf{Employing the complete form of the Killing tensor $K^{1,2,3}_{\mu \nu}$, conversely, provides no insight into this implied symmetry since both $t$ and $p$ are nullified.} 
\end{remark}

\section{The canonical forms of Killing tensor}\label{section 5}

The preservation of geometry of spacetime through a transformation reveals the existence of symmetries. These symmetries leave invariant the elements which characterize the geometry of the manifold, the metric tensor and the action apparently. Thus, the consideration of symmetries in the resolution process of the Einstein's equations is indispensable since the non-linear character of the equations of gravity obligates us to do so, in order to obtain exact solutions of EFE. 

In the following sections we present the definition of Killing tensor, the Killing equation their correlation to hidden symmetries. Afterwards, we demonstrate the order of arguments that were used by Churchill resulting in to the classification of the canonical forms of the components of a symmetric linear vector function or equivalently with an abstract symmetric tensor of valence 2.

\subsection{Killing Tensor}

The geodesic flow is a Hamiltonian system on the cotangent bundle 

\begin{equation}
    H = \frac{1}{2}g^{\mu \nu}p_\mu p_\nu
\end{equation}
where $p_\mu$ are the coordinates on the cotangent spaces or equivalently the canonical momenta of an observer. Then, an integral of motion could be defined as follows \cite{kruglikov2016geodesic}, \cite{sommers1973killing}.

\begin{equation}
 \{ \boldsymbol{\mathcal{K}} , H \} \equiv 0 \hspace{1cm} \rightarrow \hspace{1cm} \frac{\partial H}{\partial x^\mu} \frac{\partial \boldsymbol{\mathcal{K}}}{\partial p_\mu} - \frac{\partial \boldsymbol{\mathcal{K}}}{\partial x^\mu} \frac{\partial H}{\partial p_\mu} \equiv0  
\end{equation}

The function $\boldsymbol{\mathcal{K}}$: $T^* M \rightarrow \mathbb{R}$ is called polynomial of momenta and is defined as

\begin{equation}
    \boldsymbol{\mathcal{K}}(x,p) \equiv K^\mu p_\mu + K^{\mu \nu } p_\mu p_\nu + K^{\mu \nu \sigma } p_\mu p_\nu p_\sigma + K^{\mu \nu \sigma \rho } p_\mu p_\nu p_\sigma p_\rho
\end{equation} moreover, the components of the object $\boldsymbol{\mathcal{K}}$ called St$\ddot{a}$ckel-Killing tensors and satisfy the Killing equation \cite{sadeghian2022killing}, \cite{eisenhart1934separable}. The arguments in the parenthesis in relation (27) defines an endomorphism on tangent and on cotangent bundles of a smooth manifold $M$.

 $$K_{(\mu ;  \nu)} =0$$
 $$K_{(\mu \nu ; \alpha)} =0$$
 $$K_{(\mu \nu \sigma ; \alpha)} =0$$
 $$K_{(\mu \nu \sigma \rho ; \alpha)} =0$$
Killing tensors of rank r give rise to a homogeneous constant of motion of degree r in momenta. The inhomogeneous polynomial integrals of geodesic motion can be decomposed to their homogeneous parts and also to the corresponding parts that are associated with the Killing tensors with the equivalent rank. The Killing tensor of rank 1 equals to Killing vector and generates continuous symmetry transformations, these symmetries are called explicit \cite{garfinkle2010killing}. The symmetries that correspond to higher-order ranks of momenta associated with Killing tensors of rank ($r>1$) are called hidden symmetries \cite{frolov2017black}. 
The investigation of special objects such Killing tensor or Killing-Yano is basically a devilish way to peep into the phase space searching for hidden symmetries \cite{krtouvs2007killing}. Indeed, the explicit symmetries in a Hamiltonian system always could be ``dragged up" instead of hidden symmetries. These kinds of symmetries emerge during the study of the dynamics of a system featuring the conserved quantities of the system or one-parameter isometries which is equivalent with the admission of existence of Killing vectors.

As Eisenhart \cite{eisenhart1934separable} and Kalnins-Miller \cite{kalnins1980killing}, \cite{kalnins1981killing}, \cite{kalnins1983conformal} showed, the geodesic separation is correlated with the existence of Killing vectors and Killing tensors of order two \cite{benenti2016separability}. Indeed, there is a \textit{bizarre} relation between the structure of separated metrics with the structure of its characteristic Killing tensor. Benenti and Francaviglia \cite{benenti1979remarks} present a certain example where the additional information about the metric tensor serve as a catalyst in order to obtain the structure of the Killing tensor.

After this brief review about the Killing tensor let us proceed with its definition: \textbf{Any symmetric tensor of order 2 whose the symmetric part of his covariant derivative vanishes is called a Killing tensor}.
 $$K_{(\mu \nu ; \alpha)} =0$$
The \textit{trivial} Killing tensor is the metric tensor $g_{\mu \nu}$ where its existence indicates the conservation of the rest mass of a moving particle in Hamiltonian systems. The Hamiltonian is a conserved quantity of the problem since it is correlated with the conserved rest mass.

\begin{equation}\mathcal{H} = \frac{\bar{m}^2}{2} =  \frac{1}{2}g_{\mu \nu} u^\mu u^\nu \end{equation}
At last, the usage of canonical forms of a Killing tensor could be proved fruitful since can be used as a \textit{starter culture} in order to discover spacetimes with hidden symmetries.

\vspace{0.3cm}

\subsection{Canonical forms }
\vspace{0.4cm}

Obtaining the canonical forms of a symmetric 2nd-rank tensor can be a challenging task when approached algebraically.  In general, the canonical forms contain the minimum number of independent scalars which compose the eigenvalues during the diagonalization \cite{churchill1932canonical}. The equation for diagonalization in any 2-rank tensor takes the following form.

\begin{equation}\label{diag}
{K^\mu}_\nu x^\nu  - \lambda x^\mu= 0
\end{equation}
In this manner, it becomes clear that the operation of our tensor or the components of our linear vector function on a vector leaves it unchanged. In this context, it is known that every 2-rank symmetric tensor defines a linear mapping that transforms a vector $\textbf{x}$ into another vector $\textbf{v}$ unless the vector is an eigenvector. Equivalently, when a linear vector function satifies $K(\textbf{x}) = \lambda \textbf{x}$ ($\mu$ is scalar) then the direction of $\textbf{k}$ is called invariant under the action of the function $K$. We should denote that we are searching for these kinds of directions, which, of course, do not alter the quadratic form of the metric. 

\begin{equation}
   \textbf{x} \cdot \textbf{x} = x^\mu g_{\mu \nu} x^\nu = 2( x^1 x^2 - x^3 x^4)
\end{equation}
In our formalism, the norm of a vector must remain invariant under the action of the linear vector function $K$. Algebraically, the result is that a symmetric matrix $K$ is reducible under an orthogonal transformation with a matrix P to a canonical form $P K P^{-1}$ in which all non-diagonal elements are zero. 

A similar analysis can also be found in Landau-Lifshitz's book on p. 271 regarding the stress-energy-momentum tensor \cite{landau2013classical}. In this section, the authors mention that this procedure is, in fact, the application of the ``Principal Axis Theorem" or ``Spectral Theorem" for a matrix. However, it appears that the authors apply diagonalization to a symmetric tensor in a covariant form. This approach is not correct in general. Diagonalization should be applied to a tensor in a mixed tensor form, as shown in equation (\ref{diag}). The latter is always valid because ``\textit{a term proportional to the metric merely shifts all eigenvalues by the same amount}" as denoted in \cite{stephani2009exact}.

\subsubsection{The presence of null vectors within invariable planes}

In line with the work of Churchill, we obtained the canonical forms of Killing tensor \cite{churchill1932canonical}. We advise the reader to consult not only Churchill's and Rainich's work \cite{rainich1925electrodynamics} but also the work of Landau and Lifshitz (p. 271) \cite{landau2013classical} for a better understanding of the procedure. 

In this section, we provide the corresponding relations regarding our formalism, but we do not present the exact same steps as were conducted in \cite{churchill1932canonical}. Churchill's study serves as a sequel to that of Rainich's concerning a geometrical approach to find the orthogonal directions that determined by the linear symmetric vector function. Both of these works were conducted in a pseudo-Euclidean spacetime in the framework of standard formalism with signature $(-,+,+,+)$. 

In this section we focus on the corresponding components of the symmetric linear vector function $K$ which were defined by the way the function transforms the coordinate vectors and they equivalently corresponds to the components of the abstract mixed Killing tensor of second rank $K{_\mu}^\nu$ in the framework of Newman-Penrose formalism with signature $(+,-,-,-)$. 
 
$$K(\boldsymbol{i}) = K{_1}^1 \boldsymbol{i} +K{_1}^2 \boldsymbol{j} +K{_1}^3 \boldsymbol{k} +K{_1}^4 \boldsymbol{l}  $$
\begin{equation}K(\boldsymbol{j}) = K{_2}^1 \boldsymbol{i} +K{_2}^2 \boldsymbol{j} +K{_2}^3 \boldsymbol{k} +K{_2}^4 \boldsymbol{l}  \end{equation}
$$K(\boldsymbol{k}) = K{_3}^1 \boldsymbol{i} +K{_3}^2 \boldsymbol{j} +K{_3}^3 \boldsymbol{k} +K{_3}^4 \boldsymbol{l}  $$
$$K(\boldsymbol{l}) = K{_4}^1 \boldsymbol{i} +K{_4}^2 \boldsymbol{j} +K{_4}^3 \boldsymbol{k} +K{_4}^4 \boldsymbol{l} $$

It should be noted that due to the symmetry properties (for example $K(\boldsymbol{i})\boldsymbol{j} = K(\boldsymbol{j})\boldsymbol{i}$, etc. ) there are simplifications to the components of our linear symmetric vector function $K$ such as $K{_1}^2 = - K{_2}^1$. Next, we present Churchill's unitary basis vectors with respect to the null tetrads. The first basis vector behaves as a timelike vector (with a negative squared norm), while the rest behave as spacelike vectors (with positive squared norms). The latter can be verified using the orthogonality properties (see relations \ref{orthog prop}). 

$$\boldsymbol{i} \equiv \frac{l_\mu-n_\mu}{\sqrt{2}}dx^\mu $$
\begin{equation}\label{Churchill}\boldsymbol{j} \equiv \frac{l_\mu+n_\mu}{\sqrt{2}}dx^\mu \end{equation}
$$\boldsymbol{k} \equiv i\frac{m_\mu-\bar{m}_\mu}{\sqrt{2}}dx^\mu $$
$$\boldsymbol{l} \equiv \frac{m_\mu+\bar{m}_\mu}{\sqrt{2}}dx^\mu $$

Churchill initiated his study based on the theorem of Rainich who studied the anti-symmetricity of the electromagnetic tensor and he postulated that: \textit{``It is known that every linear vector function in four-dimensional space has at least one invariable plane"} \cite{rainich1925electrodynamics}. Afterwards, in a Lorentzian spacetime there are 3 kinds of 2 dimensional planes. The first case is considered as a \textbf{singular case} and the last two cases are considered as \textbf{non-singular cases} which are also be studied by Landau and Lifshitz in p. 271 \cite{landau2013classical}.

\begin{enumerate}
    \item A plane with a single null vector, like $\boldsymbol{i}+\boldsymbol{j}, \boldsymbol{k}$ plane (Singular case).
    \item A plane without null vectors, like $\boldsymbol{k}, \boldsymbol{l}$ plane (Non-singular case).
    \item A plane with 2 null vectors, like $\boldsymbol{i}, \boldsymbol{j}$ plane (Non-singular case).
\end{enumerate}

 \subsubsection{$K^0$ canonical form}
We start our technical discussion with the singular case \textbf{(Case 0} in our study) following the procedure of page 790 in \cite{churchill1932canonical}. As postulated an examination of $\boldsymbol{i}+\boldsymbol{j}, \boldsymbol{k}$ plane and $\boldsymbol{i}+\boldsymbol{j}, \boldsymbol{l}$ plane reveals that ``\textit{there is always a spacelike direction with eigenvalue $\omega_4$}". In connection to this, Churchill defines this invariable direction as a new vector $\boldsymbol{l}$ where
$$K(\boldsymbol{l}) = \omega_4 \boldsymbol{l}$$

In the remaining 3 dimensional space which is perpendicular to $\boldsymbol{l}$ direction and satisfies the existence of a single null vector there is a function $\phi(\boldsymbol{x}) \equiv K(\boldsymbol{x}) - \lambda_1 \boldsymbol{x}$. The $\phi(\boldsymbol{x})$ has the same invariable directions as $K(\boldsymbol{x})$ but different multipliers. Furthermore it transforms every vector $\boldsymbol{x}$ that belongs to $\boldsymbol{i}, \boldsymbol{j}, \boldsymbol{k}$ space into a vector of $\boldsymbol{i} +\boldsymbol{j}, \boldsymbol{k}$ plane, also it transforms every vector in this plane into a null vector aligned with $\boldsymbol{i} +\boldsymbol{j}$ direction. Finally, after the determination of $\phi(\boldsymbol{x})$ function, Churchill results in function $K(\boldsymbol{x})$ where $\lambda_1, \omega_4$  are the only multipliers of the invariable directions of $\boldsymbol{i} +\boldsymbol{j}, \boldsymbol{l}$ accordingly.

We present the canonical form obtained by Churchill (relation (21) in \cite{churchill1932canonical}), with a single difference in the second term of the first two relations, which arises due to metric signature. The last statement can also be verified through the symmetry property $K(\boldsymbol{i}) \boldsymbol{k} = K(\boldsymbol{k})\boldsymbol{i}$. 

$$K(\boldsymbol{i})= \lambda_1 \boldsymbol{i} + \boldsymbol{k} $$
$$K(\boldsymbol{j})= \lambda_1 \boldsymbol{j} - \boldsymbol{k} $$
$$K(\boldsymbol{k})= \boldsymbol{i} +\boldsymbol{j} +\lambda_1\boldsymbol{k}$$
$$K(\boldsymbol{l}) = \omega_4 \boldsymbol{l}$$

If we try to express the above relations with respect to null tetrads\footnote{For the sake of simplification, we present these relations without indices.} we get the following relations 
$$K(n) = \lambda_1 n -i ((-\bar{m})- (-m)) $$ 
\begin{equation}K(l) = \lambda_1 l\end{equation}
$$K(-\bar{m})= -il -\lambda_2(-\bar{m}) -\lambda_7 (-m) $$
$$K(-m)= +il -\lambda_7(-\bar{m}) -\lambda_2 (-m) $$
where we define that $-\lambda_2 \equiv \frac{\omega_4+\lambda_1}{2}$ and $-\lambda_7 \equiv \frac{\omega_4-\lambda_1}{2}$. In the matrix representation we present the mixed Killing form and with the usage of the metric we can obtain the covariant Killing form apparently.
\begin{equation} {K{^0}}{_{\mu}}^\nu =  \begin{pmatrix}
\lambda_1 &0 & \bar{p} & p \\
  0   & \lambda_1 &  0 & 0\\
0  &   -p &  \lambda_7   &\lambda_2 \\
0  &   -\bar{p} & \lambda_2 & \lambda_7 
\end{pmatrix}
\Longleftrightarrow
 {K{^0}}_{\mu \nu} =  \begin{pmatrix}
0 & \lambda_1   & -p & -\bar{p} \\
 \lambda_1 & 0 & 0 & 0\\
-p & 0&  \lambda_7   &\lambda_2 \\
-\bar{p} & 0 & \lambda_2 & \lambda_7 
\end{pmatrix} ; \hspace{0.3cm} p=-\bar{p} = \pm i
\end{equation}

\vspace{0.2cm}

Diagonalizing the $K^0{_\mu}^\nu$ form one can easily find that this form has one triple eigenvalue, namely $\lambda_1= -(\lambda_2 -\lambda_7)$ which is placed in a 3-dimensional Jordan canonical form. In addition, the other eigenvalue described by the form $\omega_4 =  -(\lambda_2+\lambda_7)$.

\subsubsection{$K^1$, $K^2$, $K^3$ canonical forms}
In this section we are going to provide the general form of the mixed Killing tensor concerning all three cases. Afterwards, we will operate a classification based on the sign of the discriminant of the equation (\ref{roots}) that emerges during the diagonalization.

Starting by the assumption of existence of zero null vectors in the invariable plane of $K(\boldsymbol{x})$ we can find new $\boldsymbol{k}$ and $\boldsymbol{l}$ vectors lie in this invariable plane and transform them to invariable directions. Building upon this, the new $\boldsymbol{i}, \boldsymbol{j}$ plane is also invariable and perpendicular to the former. 

On the other hand initiating by the assumption of existence of two null vectors in the invariable plane of $K(\boldsymbol{x})$ we can find new $\boldsymbol{i}, \boldsymbol{j}$ which can be made to fall into this invariable plane. Using the symmetry property we can find new $\boldsymbol{k}$ and $\boldsymbol{l}$ vectors which are lie to an invariable plane with zero null vectors, therefore in this plane there are two invariable directions as in the previous non-singular case.

Hence, both cases yield the same 2-rank mixed Killing tensor, with the main difference being the number of null vectors in the $\boldsymbol{i}, \boldsymbol{j}$ plane. 

$$K(\boldsymbol{i})= \alpha \boldsymbol{i} + \beta \boldsymbol{j} $$
\begin{equation}\label{directions} K(\boldsymbol{j})= -\beta \boldsymbol{i} + \gamma\boldsymbol{j}\end{equation} 
$$K(\boldsymbol{k})=  \omega_3   \boldsymbol{k}$$
$$K(\boldsymbol{l}) = \omega_4   \boldsymbol{l}$$
An examination of this plane will distinct the different cases of the canonical forms. A vector $\boldsymbol{x} = x^1 \boldsymbol{i} + x^2 \boldsymbol{j}$ belongs to invariable direction if

$$K(x^1 \boldsymbol{i} + x^2 \boldsymbol{j})= \omega (x^1 \boldsymbol{i} + x^2 \boldsymbol{j})$$
Subsequently, we take 
$$x^1(\alpha - \omega) - x^2 \beta=0$$
$$x^1 \beta + x^2 (\gamma - \omega)=0$$
By eliminating $x^1$ and $x^2$, we investigate solutions where neither of these variables are zero. 
\small
\begin{equation}\label{roots}
    \omega^2 -\omega(\alpha+\gamma) +\beta^2 +\alpha\gamma = 0 \Longrightarrow  \omega_{1,2} = \frac{(\alpha +\gamma) \pm\sqrt{(\alpha+\gamma)^2 -4(\beta^2 +\alpha \gamma)}}{2}
\end{equation}
\normalsize
This equation yield two solutions $\omega_1$ and $\omega_2$. Within this scheme, three cases arise from the examination of the discriminant. The first case (\textbf{Case 1}) emerges when the discriminant is set to zero, resulting in a double root with $\omega_1=\omega_2=\frac{\alpha +\gamma}{2}$. In contrast to Churchill's study during this examination we did not consider the cases where 1) $\beta=0$ as it is a special case of \textbf{Case 3} and 2) $\beta\neq0$, where $\omega_1=\omega_2 = 0$. However, in \textbf{Case 1} we obtained two different forms that result in the same diagonalized form.

This is evident if we use the relation (14) from \cite{churchill1932canonical}, where $\psi_1 = 2\sqrt{\psi_2} = \omega_1$.    
\begin{equation}  
{K{^{1a}}}{_{\mu}}^\nu = \begin{pmatrix}
0 & \lambda_1   & 0 & 0\\
 -\lambda_1 & \lambda_0 & 0 & 0\\
0 & 0&  \omega_3   &0 \\
0 & 0 & 0 & \omega_4   
\end{pmatrix}  \hspace{0.1cm},
{K{^{1b}}}{_{\mu}}^\nu = \begin{pmatrix}
 \lambda_0 &\lambda_1   & 0 & 0\\
 -\lambda_1 & 0 & 0 & 0\\
0 & 0&  \omega_3   &0 \\
0 & 0 & 0 & \omega_4  
\end{pmatrix} 
\end{equation}
The transformation in null basis vectors using the relations (\ref{Churchill}) and setting $\omega_1 = \lambda_1$, $\lambda_0 = \pm 2\omega_1$, $-\lambda_2 \equiv \omega_4+\omega_3$ and $-\lambda_7 \equiv \omega_4-\omega_3$ \footnote{It should be noted that $\lambda_0$ does not contribute to the calculations and one could verify that for any scalar value of $\lambda_0$ the diagonalization results in a 2 dimensional Jordan block with a double eigenvalue $\lambda_1$ in the upper-left block.} yields the two canonical forms with one double eigenvalue $\lambda_1$ and two single eigenvalues $-(\lambda_2\pm\lambda_7)$.
 
\begin{equation}  
{K{^{1a}}}_{\mu \nu} = \begin{pmatrix}
0 & \lambda_1   & 0 & 0\\
 \lambda_1 & \lambda_0 & 0 & 0\\
0 & 0&  \lambda_7   &\lambda_2 \\
0 & 0 & \lambda_2 & \lambda_7   
\end{pmatrix}  \hspace{0.1cm},
{K{^{1b}}}_{\mu \nu} = \begin{pmatrix}
\lambda_0 & \lambda_1   & 0 & 0\\
 \lambda_1 & 0 & 0 & 0\\
0 & 0&  \lambda_7   &\lambda_2 \\
0 & 0 & \lambda_2 & \lambda_7   
\end{pmatrix} 
\end{equation}
This case produces two forms but considered as one by us. The reason is based on the concept of the symmetrical null tetrad frame which introduced by Debever in \cite{debever1984nouvelle}. This symmetry refers to the interchanges between the tetrads $n^\mu \leftrightarrow l^\mu$  and $m^\mu \leftrightarrow \bar{m}^\mu$.  

The next case is the most general one, as it contains two distinct real eigenvalues and corresponds to \textbf{Case 2}. This case arises when the discriminant is positive. Hence, there are 4 invariable directions, and the mixed Killing tensor appears in a diagonal form with eigenvalues $\omega_1, \omega_2, \omega_3, \omega_4$. Using again the relations (\ref{Churchill}) and setting $\lambda_1 \equiv \omega_2+\omega_1$ and $\lambda_0 \equiv \omega_2-\omega_1$, $-\lambda_2 \equiv \omega_4+\omega_3$ and $-\lambda_7 \equiv \omega_4-\omega_3$

\begin{equation}
{K{^2}}{_\mu}^\nu = \begin{pmatrix}
\lambda_1 & \lambda_0   & 0 & 0\\
 \lambda_0 & \lambda_1 & 0 & 0\\
0 & 0&  -\lambda_2   &-\lambda_7 \\
0 & 0 & -\lambda_7 & -\lambda_2   
\end{pmatrix}  \hspace{0.1cm} \Longleftrightarrow 
 {K{^2}}_{\mu \nu} = \begin{pmatrix}
\lambda_0 & \lambda_1   & 0 & 0\\
 \lambda_1 & \lambda_0 & 0 & 0\\
0 & 0&  \lambda_7   &\lambda_2 \\
0 & 0 & \lambda_2 & \lambda_7   
\end{pmatrix}
\end{equation}

Finally, \textbf{Case 3} arises when the discriminant takes negative values. In this scenario, there are no invariable directions within the $\boldsymbol{i}, \boldsymbol{j}$ plane. Consequently, we are unable to find more useful directions and must work with the available one. Taking advantage of relations ($\ref{directions}$) and combine them with relations ($\ref{Churchill}$) we  get the following

\begin{equation}
    K(n) = \frac{\alpha+\gamma}{2} n + \frac{\alpha -\gamma -2\beta}{2} l
\end{equation}
\begin{equation}
    K(l) = -\frac{\alpha -\gamma -2\beta}{2} n +\frac{\alpha +\gamma }{2}l 
\end{equation}
In this point we define $\lambda_1 \equiv \frac{\alpha+\gamma}{2}$, $\lambda_0 \equiv \frac{\alpha -\gamma -2\beta}{2}$, $-\lambda_2 \equiv \omega_4+\omega_3$ and $-\lambda_7 \equiv \omega_4-\omega_3$

It should be noted that in this manner the existence of one timelike and one spacelike eigenvector is satisfied since it serves as a constraint to us. Generally speaking this is the reason why in a pseduo-Euclidean spacetime we cannot have two pairs of complex conjugate eigenvalues \cite{landau2013classical}. After the definitions above we result in the mixed tensor form and with the usage of metric of our formalism we obtain the covariant one.

\begin{equation}
{K{^3}}{_\mu}^\nu = \begin{pmatrix}
\lambda_1 & \lambda_0   & 0 & 0\\
 -\lambda_0 & \lambda_1 & 0 & 0\\
0 & 0&  -\lambda_2   &-\lambda_7 \\
0 & 0 & -\lambda_7 & -\lambda_2   
\end{pmatrix}  \hspace{0.1cm} \Longleftrightarrow
 {K{^3}}_{\mu \nu} = \begin{pmatrix}
\lambda_0 & \lambda_1   & 0 & 0\\
 \lambda_1 & -\lambda_0 & 0 & 0\\
0 & 0&  \lambda_7   &\lambda_2 \\
0 & 0 & \lambda_2 & \lambda_7   
\end{pmatrix} \end{equation}
\vspace{0.2cm}

It should be noted that the only difference between $K^1$, $K^2$ and $K^3$ could be described via a factor $q\equiv 0,\pm1$. 

\subsubsection{The diagonalized form of the Canonical forms}\label{diagonalization}

We present the diagronalized canonical forms of the Killing tensor. 

\footnotesize

\begin{equation}\tag{Case 0}
{K^0_{\mu}}^\nu =
\begin{pmatrix}
\lambda_1 & 1 & 0 & 0\\
0&\lambda_1 & 1 & 0\\
0 & 0&  \lambda_1&0 \\
0 & 0 & 0&-(\lambda_2 + \lambda_7)   
\end{pmatrix} ;
\hspace{0.1cm} 
\lambda_1 = - (\lambda_2 - \lambda_7)
\end{equation}

\begin{equation}\tag{Case 1}
{K^1_{\mu}}^\nu = \begin{pmatrix}
\lambda_1 & 1 & 0 & 0\\
0&\lambda_1 & 0 & 0\\
0 & 0&  -(\lambda_2 +\lambda_7)&0 \\
0 & 0 & 0&-(\lambda_2 - \lambda_7)   
\end{pmatrix}
\end{equation}

\begin{equation}\tag{Case 2}
{K^2_{\mu}}^\nu = \begin{pmatrix}
\lambda_1 +\lambda_0  & 0 & 0\\
0&\lambda_1- \lambda_0 & 0 & 0\\
0 & 0&  -(\lambda_2 +\lambda_7)&0 \\
0 & 0 & 0&-(\lambda_2 - \lambda_7)   
\end{pmatrix}
\end{equation}

\begin{equation}\tag{Case 3}
 {K^3_{\mu}}^\nu = \begin{pmatrix}
\lambda_1 +i\lambda_0  & 0 & 0\\
0&\lambda_1- i\lambda_0 & 0 & 0\\
0 & 0&  -(\lambda_2 +\lambda_7)&0 \\
0 & 0 & 0&-(\lambda_2 - \lambda_7) \end{pmatrix}
\end{equation}

\vspace{0.3cm}
\normalsize

In conclusion, the canonical forms encompass all the needed information of an arbitrary Killing tensor, categorized according to the number of its eigenvalues and their multiplicities. It is also obvious that the four canonical forms are generalizations of the Killing form with two double eigenvalues (\textit{Paradigm}). The solution extraction through the canonical forms scoping to elevate the usage of Killing tensor. 

\vspace{1cm}

\section{Solutions of $K^0_{\mu \nu}$ form}\label{section 6}
We initiate the second part of this work by the study of the first canonical form $K^0_{\mu \nu}$ form. 
\
\begin{multline}K^0_{\mu \nu} =  \lambda_1(l_\mu n_\nu + n_\mu l_\nu ) - p(n_\mu m_\nu +m_\mu n_\nu) -\bar{p}(n_\mu \bar{m}_\nu +\bar{m}_\mu n_\nu ) \\
+ \lambda_7(m_\mu m_\nu  +\bar{m}_\mu \bar{m}_\nu)+ \lambda_2 ( m_\mu \bar{m}_\nu +\bar{m}_\mu m_\nu) ; \hspace{0.2cm} p=-\bar{p} = \pm i\end{multline}
\normalsize

It will be proved really helpful to define the factor $Q$, which is a real quantity since it depends on the elements of each canonical form. In the next chapters the corresponding factor will be redefined for each form differently. The factor $Q$ for $K^0_{\mu \nu}$ is defined by

\begin{equation}
    Q\equiv \frac{\lambda_7}{\lambda_1+\lambda_2}
\end{equation}

The relations (\ref{nderivative})-(\ref{mderivative}) are indispensable in order to extract any useful relation by the Killing equation $K^0_{(\mu \nu;\alpha)} = 0$. The annihilation of the symmetrised covariant derivation of our canonical form yields the following relations. 

\begin{equation}\label{eq38}
    \nu(\lambda_1 +\lambda_2) +\bar\nu \lambda_7 = 0
\end{equation}
\begin{equation}
    \kappa p+ \bar\kappa \bar{p} = 0
\end{equation}
\begin{equation}
    \bar\kappa (\lambda_1 +\lambda_2) +\kappa \lambda_7 +(\rho - \bar\epsilon)p +\bar\sigma\bar{p} = 0
\end{equation}
\begin{equation}
    \kappa(\lambda_1 + \lambda_2) +\bar\kappa \lambda_7 +(\bar\rho - \epsilon)p+\sigma\bar{p}
    \end{equation}

\begin{equation}
    D\lambda_1 +(\tau-\bar\pi)p +(\bar\tau-\pi)\bar{p} = 0
\end{equation}
\begin{equation}
    \Delta \lambda_1 -\bar\nu p - \nu \bar{p} = 0
\end{equation}
\begin{equation}
-\delta\lambda_1 +(\bar\pi - \tau)(\lambda_1 +\lambda_2) +(\pi -\bar\tau)\lambda_7 +\bar\lambda p +(\mu +2\gamma)\bar{p} = 0
\end{equation}
\begin{equation}
-\bar\delta\lambda_1 +(\pi - \bar\tau)(\lambda_1 +\lambda_2) +(\bar\pi -\tau)\lambda_7 +(\bar\mu +2\bar\gamma)p+\lambda\bar{p} = 0 
\end{equation}

\begin{equation}
    D\lambda_2 +(\rho +\bar\rho)(\lambda_1+\lambda_2) +(\sigma +\bar\sigma)\lambda_7 -(\bar\pi +2\bar\alpha)p - (\pi +2\alpha)\bar{p} = 0
\end{equation}
\begin{equation}
    \Delta\lambda_2 - (\mu +\bar\mu)(\lambda_1 +\lambda_2) -(\lambda+\bar\lambda)(\lambda_1+\bar\lambda) - \bar\nu p -\nu \bar{p} = 0
\end{equation}
\begin{equation}
    -2\delta \lambda_2 -\bar\delta\lambda_7 +2(\alpha - \bar\beta)\lambda_7 +2\bar\lambda p+2(\mu +\bar\mu)\bar{p}= 0 
\end{equation}
\begin{equation}
    -2\bar\delta\lambda_2 - \delta\lambda_7+2(\bar\alpha-\beta)\lambda_7 +2(\mu +\bar\mu)p +2\lambda\bar{p} = 0
\end{equation}

\begin{equation}
D\lambda_7 +2\bar\sigma(\lambda_1+\lambda_2) +2(\rho+\epsilon-\bar\epsilon)\lambda_7
-2p(\pi +2\bar\beta) = 0\end{equation}
\begin{equation}
    D\lambda_7 +2\sigma(\lambda_1 +\lambda_2) +2(\bar\rho -\epsilon +\bar\epsilon)\lambda_7 -2\bar{p}(\bar\pi +2\beta) = 0
\end{equation}
\begin{equation}
    \Delta\lambda_7 -2\lambda(\lambda_1+\lambda_2)+2(\gamma -\bar\gamma -\bar\mu)\lambda_7 -2\nu p=0
\end{equation}
\begin{equation}
    \Delta\lambda_7 -2\bar\lambda (\lambda_1+\lambda_2)+ 2(\bar\gamma - \gamma -\mu)\lambda_7-2\bar\nu \bar{p} = 0
\end{equation}
\begin{equation}
    -\delta\lambda_7 -2(\bar\alpha -\beta)\lambda_7+2\bar{p}\bar\lambda=0
\end{equation}
\begin{equation}
-\bar\delta\lambda_7 -2(\alpha -\bar\beta)\lambda_7
+2p\lambda =0\end{equation}

We ought to mark that the annihilation of the real and the imaginary part of the equation (\ref{eq38}) yields the following two possibilities.

$$(\nu+\bar\nu)(\lambda_1 +\lambda_2 +\lambda_7) = 0 \Leftrightarrow (\nu+\bar\nu)(Q+1) = 0 $$
$$(\nu -\bar\nu)(\lambda_1 +\lambda_2 -\lambda_7) = 0 \Leftrightarrow(\nu-\bar\nu)(Q-1) = 0$$

We are already aware that $Q+1=0$. Then, the equations regardless the value of the factor $p$ yield 

\begin{equation}\label{eq56}
    \epsilon - \bar\epsilon = \kappa-\bar\kappa = \nu = \gamma = \mu = \lambda = 0
\end{equation}
\begin{equation}
    \pi - \bar\pi +2(\alpha - \bar\alpha) = 0
\end{equation}
\begin{equation}
    \pi +2\bar\beta = 0
\end{equation}
\begin{equation}
    \rho - \bar\sigma = \epsilon
\end{equation}
\begin{equation}\label{eq61}
    \pi - \bar\pi + \tau - \bar\tau = 0
\end{equation}

\begin{equation}
D\lambda_1 = \Delta \lambda_1 = \delta \lambda_1 = 0    
\end{equation}
\begin{equation}
    D\lambda_2 = 2(\lambda_1 +\lambda_2)(\sigma - \bar\rho)
\end{equation}
\begin{equation}
    \Delta\lambda_2 =0
\end{equation}
\begin{equation}
    \delta \lambda_2 = 2(\lambda_1+\lambda_2)(\beta - \bar\alpha)
\end{equation}
\begin{equation}
    D\lambda_7 = 2\lambda_7(\sigma - \bar\rho)
\end{equation}
\begin{equation}
    \Delta \lambda_7 = 0
\end{equation}
\begin{equation}
    \delta\lambda_7 = 2\lambda_7 (\beta - \bar\alpha)
\end{equation}

Since we obtained the derivatives of $\lambda$'s we are able now to derive the integrability conditions of $K^0_{\mu \nu}$ for the first case. Based on the equation $Q+1 = \lambda_1 +\lambda_2 +\lambda_7 = 0$ we can easily check that 

\begin{equation}
    \frac{D(\lambda_1+\lambda_2)}{\lambda_1 +\lambda_2} = \frac{D\lambda_7}{\lambda_7} = 2(\sigma-\bar\rho)
\end{equation}
\begin{equation}
     \frac{\delta(\lambda_1+\lambda_2)}{\lambda_1 +\lambda_2} = \frac{\delta\lambda_7}{\lambda_7} = 2(\beta - \bar\alpha)
\end{equation}

\subsection{Integrability conditions of $K^0_{\mu \nu}$}

The integrability conditions for $K^0_{\mu \nu}$ form are given as follows.

\begin{equation}\tag{CR1:$\lambda_7$}
\Delta(\sigma - \bar\rho) = (\alpha -\bar\beta)(\pi +\bar\pi +\tau+\bar\tau)
\end{equation}
\begin{equation}\tag{CR2:$\lambda_7$}2\delta(\sigma-\bar\rho) +2D(\alpha-\bar\beta) =(\sigma-\bar\rho)(\bar\alpha +\beta -\bar\pi) -(\alpha-\bar\beta)(\sigma+\bar\rho)
\end{equation}
\begin{equation}\tag{CR3:$\lambda_7$}
    \Delta(\alpha-\bar\beta) =0
\end{equation}
\begin{equation}\tag{CR4:$\lambda_7$}
    (\delta - \bar\delta)(\alpha-\bar\beta) = 0
\end{equation}

The relations (\ref{eq56})-(\ref{eq61}) and the integrability conditions provide us with further information crystallizes the usefulness of the entanglement of the Killing tensor. The latter along with the Newman-Penrose equations and the Bianchi identities form a solvable-overdetermined system of equations. 

\small
\begin{equation}\tag{a} D \rho  - \bar{\delta} \kappa = {\rho}(\rho+2\epsilon) +\sigma \bar\sigma   - \kappa \left[2(\alpha +\bar{\beta}) + (\alpha - \bar{\beta}) - \pi +\tau \right] +\Phi_{00}\end{equation}
\begin{equation}\tag{b}\label{b} \delta \kappa - D\sigma = - (\rho +\bar\rho +2\epsilon)\sigma +\kappa \left[ \tau - \bar{\pi} +2(\bar{\alpha} +\beta) - (\bar{\alpha} - \beta) \right] - \Psi _o  \end{equation}
\begin{equation}\tag{c} D\tau - \Delta \kappa = \rho(\tau + \bar{\pi}) + \sigma(\pi +\bar\tau) + \Psi_1  \end{equation}
\begin{equation}\tag{i}\label{i}  - \Delta \pi = \Psi_3  \end{equation}
\begin{equation}\tag{g} \bar{\delta} \pi  = - \pi(\pi + \alpha - \bar{\beta}) \end{equation}
\begin{equation}\tag{p}\label{p} \delta \tau -\Delta\sigma =  \tau (\tau - \bar{\alpha} + \beta) \end{equation}
\begin{equation}\tag{h} - \delta\pi =  \pi(\bar{\pi} - \bar{\alpha} +\beta)+ \Psi_2 + 2\Lambda \end{equation}
\begin{equation}\tag{q} \Delta \rho - \bar{\delta} \tau = - \tau(\bar{\tau} + \alpha - \bar{\beta}) - \Psi_2 - 2\Lambda\end{equation}
\begin{equation}\tag{k}\label{k} \delta\rho -\bar\delta\sigma= \rho(\bar{\alpha} + \beta) - \sigma(3\alpha-\bar\beta) +\tau(\rho-\bar{\rho}) - \Psi_1  \end{equation}
\begin{equation}\tag{m}\label{m} 0 = \Psi_3 \end{equation}
\begin{equation}\tag{d} D\alpha - \bar{\delta} \epsilon = \alpha(\rho -\epsilon) +\beta\bar\sigma- \bar{\beta}\epsilon + \pi (\epsilon + \rho) \end{equation}
\begin{equation}\tag{e}\label{e} D \beta - \delta{\epsilon} = \sigma(\alpha+\pi) +\beta(\bar{\rho} - \bar{\epsilon})  -\epsilon(\bar{\alpha} - \bar{\pi}) + \Psi_1\end{equation} 
\begin{equation}\tag{r} \Delta \alpha = \Psi_3 \end{equation}
\begin{equation}\tag{o}\label{o} \Delta \beta =0\end{equation}
\begin{equation}\tag{l}\label{l} \delta \alpha - \bar{\delta}\beta = \alpha (\bar{\alpha} - \beta) - \beta(\alpha - \bar{\beta})-\Psi_2 + \Lambda \end{equation}
\begin{equation}\tag{f}\label{f} - \Delta \epsilon = \alpha(\tau + \bar{\pi}) + \beta( \bar{\tau} + \pi)  + \Psi_2 - \Lambda + \tau \pi\end{equation}
\begin{equation}\tag{j} 0 = \Psi_4\end{equation}
\normalsize
The Bianchi identities with $\Psi_3 = \Psi_4 =0$ is given below.

\small
\begin{equation}\tag{I}\label{I} \bar{\delta} \Psi_0 - D \Psi_1   = (4\alpha - \pi )\Psi_0  - 2(2\rho +\epsilon)\Psi_1 +3\kappa \Psi_2 \end{equation}
\begin{equation}\tag{II}\label{II} \bar{\delta} \Psi_1 -D\Psi_2  =  2(\alpha - \pi)\Psi_1 -3\rho \Psi_2   \end{equation}
\begin{equation}\tag{III}\label{III} \bar{\delta}\Psi_2 = -3\pi \Psi_2\end{equation}
\begin{equation}\tag{V}\label{V} \Delta \Psi_0 - \delta \Psi_1 =  -2(2\tau +\beta)\Psi_1 +3\sigma\Psi_{2}  \end{equation}
\begin{equation}\tag{VI}\label{VI} \Delta \Psi_1 - \delta \Psi_2  = -3\tau\Psi_2 \end{equation}
\begin{equation}\tag{VII}\label{VII} \Delta \Psi_2 =0 \end{equation} 
\normalsize

\subsection{A unique Petrov type D solution}

At this point, we must ascertain that none of the remaining Weyl components are nullified. Now, we will prove that $\Psi_2$ is non-zero, thereby confirming that this solution is of Petrov type D.
The real and imaginary part of $\Psi_2$ could be obtained by the comparison of NPE (\ref{p}) with ($CR1:\lambda_7$) considering the relation $\epsilon = \rho - \bar\sigma$.

\begin{equation}
    \Psi_2 +\bar{\Psi}_2 -2\Lambda = (\alpha - \bar\beta)(\pi +\bar\pi +\tau+\bar\tau) +2 \pi \bar\pi
\end{equation}
\begin{equation}
    \Psi_2 - \bar{\Psi}_2 = (\bar\tau-\tau)\left[2(\alpha-\bar\beta) +\pi -\bar\tau \right]
\end{equation}
We are not be able to determine if this solution is a brand new type D solution, however the main characteristics of this type D solution is inferred in the downward relations. 

\begin{equation}
    \kappa\sigma\rho  \neq 0
\end{equation}
\begin{equation}
    \Psi_4 = C_{\kappa \lambda \mu \nu} n^\kappa \bar{m}^\lambda n^\mu \bar{m}^\nu = 0 
\end{equation}
\begin{equation}
     \Psi_3 = C_{\kappa \lambda \mu \nu} n^\kappa l^\lambda n^\mu \bar{m}^\nu = 0 
\end{equation}

Furthermore, we should note that any rotation around the null tetrad frame is inapplicable due to the non-preservation of the Killing tensor during rotation. Also, considering the aforementioned relations, there is a non-geodesic ($\kappa \neq 0$), a non-diverging ($\rho \neq 0$) null congruence characterized by shear ($\sigma \neq0$) and our solution is of type D, with $\Psi_2 \Psi_0 \Psi_1 \neq 0$, and the null vector $n^\mu$ serve as a Principal Null Direction (PND).  

\vspace{0.5cm}

\begin{theorem} Assuming the existence of $K^0_{\mu \nu}$ canonical form of Killing tensor in vacuum with $\Lambda$ a Petrov type D solution with a non-geodesic, shearing and diverging null congruence $n^\mu$ emerged.
\end{theorem}

\section{Solutions of $K^1_{\mu \nu}$ form}\label{section 7}

The $K^1_{\mu \nu}$ form has been handled in a unique manner. We nullify $\lambda_7$ and set $\lambda_0$ to be a constant equal to a value represented as $q=\pm1$. This approach was chosen to achieve a similar form to that of Hauser-Malhiot (Paradigm) with the only difference being the constant $q$.

With these simplifications, we obtain a Jordan form of Killing tensor, as opposed to a diagonalized form with two double eigenvalues. Besides, it is widely known that the Jordan canonical form of a matrix embodies all the similar matrices of the family of matrices with the same eigenvalues except the ``unique" member of this family, the diagonalized member of the family \cite{strang2022introduction}. We shall proceed now with the Killing equation serves as a starting point,

\begin{equation}K^1_{\mu \nu} = qn_\mu n_\nu + \lambda_1(l_\mu n_\nu + n_\mu l_\nu )  + \lambda_2 ( m_\mu \bar{m}_\nu +\bar{m}_\mu m_\nu)\end{equation}

To acquire solutions of NPE, first, we have to consider additional mathematical conditions. In our case these represent the integrability conditions of the eigenvalues $\lambda_1$, $\lambda_2$ of our Killing tensor which come by the Killing equation. 

\begin{equation} K_{(\mu \nu ; \alpha)} = 0\end{equation}

The Killing equation provides us with the following relations. They will be used along with the CR in order to give us the IC of the eigenvalues $\lambda_1,\lambda_2$,

\begin{equation}\nu =\sigma = \lambda = 0\end{equation}
\begin{equation}\label{6.4}q (\epsilon +\bar{\epsilon}) = 0 \hspace{1cm}q\neq0 \end{equation}
\begin{equation}\label{eq80}(\lambda_1 + \lambda_2)\kappa=q(\bar{\alpha} + \beta + \bar{\pi}) \end{equation}
The directional derivatives for eigenvalues $\lambda_1$, $\lambda_2$ turned out to be the following

\begin{equation} D\lambda_1 = q (\gamma + \bar{\gamma}) \end{equation}
\begin{equation}\Delta \lambda_1 = 0 \end{equation}
\begin{equation}\delta \lambda_1 =(\lambda_1 +\lambda_2)(\bar{\pi} - \tau)\end{equation}

\begin{equation}D\lambda_2 = q(\mu + \bar{\mu}) -(\lambda_1 +\lambda_2) (\rho +\bar{\rho}) \end{equation}
\begin{equation} \Delta \lambda_2 = (\lambda_1 +\lambda_2)   (\mu + \bar{\mu})\end{equation}
\begin{equation}  \delta \lambda_2 = 0\end{equation}
The relation (\ref{eq80}) could be used defining a factor $Q$ with its directional derivatives

\begin{equation}Q \equiv \frac{q}{\lambda_1 +\lambda_2} = \frac {\kappa}{ \bar{\alpha} + \beta  + \bar{\pi}  } \end{equation}

\begin{equation}DQ = Q [ \rho + \bar{\rho} -Q  (\gamma + \bar{\gamma} + \mu + \bar{\mu}) ]\end{equation}
\begin{equation}\Delta Q = -Q  (\mu + \bar{\mu})\end{equation}
\begin{equation}\delta Q = -Q(\bar{\pi}-\tau)\end{equation}
The factor $Q$ was proved helpful in the treatment of the IC and it is a real function since it depends only from real eigenvalues.

\vspace{0.3cm}
\subsection{Integrabillity Conditions of $K^1_{\mu\nu}$ form}

IC come to surface by acting of the CR upon to the eigenvalues. Additionally, CR is the Lie bracket of the basis vectors. We choose to separate the IC in two parts using the factor $Q$. 

\vspace{0.2cm}

Integrability Conditions of Eigenvalue $\lambda_1$ 
\small
\begin{equation}\tag{$CR1:\lambda_1$}Q[\delta(\gamma+\bar{\gamma}) - (\gamma+\bar{\gamma})(\bar{\alpha} +\beta -\tau)-(\mu+\bar{\mu})(\bar{\pi} -\tau)] = D(\bar{\pi} -\tau) -(\bar{\pi} -\tau)(\rho+\bar{\rho}) - (\bar{\pi} -\tau)(2\epsilon+\bar{\rho})  \end{equation}
\begin{equation}\tag{$CR2:\lambda_1$} Q[\Delta(\gamma+\bar\gamma) - (\gamma +\bar{\gamma})^2] = 2(\tau \bar{\tau} - \pi \bar{\pi})  \end{equation}
\begin{equation}\tag{$CR3:\lambda_1$} \Delta(\bar{\pi} -\tau) -(\bar{\pi} -\tau)[(\gamma - \bar{\gamma})-(2\mu +\bar{\mu})] = 0 \end{equation}
\begin{equation}\tag{$CR4:\lambda_1$} \delta(\pi-\bar{\tau})-\bar{\delta}(\bar{\pi}-\tau) +(\bar{\pi} - \tau)(\alpha-\bar{\beta})-(\pi-\bar{\tau})(\bar{\alpha}-\beta)=Q(\gamma+\bar{\gamma})(\mu - \bar{\mu}) \end{equation}
\vspace{0.2cm}

\normalsize
Integrability Conditions of Eigenvalue $\lambda_2$ 
\small
\begin{equation}\tag{$CR1:\lambda_2$}Q[\delta(\mu+\bar{\mu}) - 2(\mu+\bar{\mu})(\bar{\alpha} +\beta)] = \delta(\rho+\bar{\rho})-(\rho+\bar{\rho})[(\bar{\alpha} +\beta-\bar{\pi})-(\bar{\pi} - \tau)]   \end{equation}
\begin{equation}\tag{$CR2:\lambda_2$} Q[\Delta(\mu+\bar{\mu}) -(\mu+\bar{\mu})[(\mu+\bar{\mu})+2(\gamma +\bar{\gamma})]] = D(\mu+\bar{\mu}) +\Delta(\rho+\bar{\rho}) - (\rho+\bar{\rho})(\gamma+\bar{\gamma})    \end{equation}
\begin{equation}\tag{$CR3:\lambda_2$} \delta(\mu+\bar{\mu})+(\mu+\bar{\mu})[(\bar{\alpha} +\beta - \tau)+(\bar{\pi}-\tau)] = 0 \end{equation}
\begin{equation}\tag{$CR4:\lambda_2$} 2(\mu \bar{\rho} - \bar{\mu}\rho) = Q(\mu+\bar{\mu})(\mu-\bar{\mu}) \end{equation}
\normalsize

\subsection{Classes of solutions}

Consider now the $K^1$ Killing form.

 $$K^1 = q \tilde{\theta}^1 \otimes \tilde{\theta}^1 +\lambda_1(\tilde{\theta}^1 \otimes \tilde{\theta}^2+\tilde{\theta}^2 \otimes \tilde{\theta}^1) + \lambda_2(\tilde{\theta}^3 \otimes \tilde{\theta}^4+\tilde{\theta}^4 \otimes \tilde{\theta}^3); \hspace{0.5cm} q= \pm 1$$
This Killing form dictates that the only non-zero rotation parameter is $t=ib$. With the usage of the general \textit{key relations} (\ref{kri})-(\ref{kriv}) of chapter 4 we get the following relations. 

 \begin{equation}\tag{i} \Psi_2 - \Lambda = \tau \pi \end{equation}
 \begin{equation}\tag{ii}\Psi_1 = \kappa \mu\end{equation}
 \begin{equation}\tag{iii}\Psi_2 - \Lambda = \mu \rho\end{equation}
\begin{equation}\tag{iv}\mu \tau = 0\end{equation}

 The rotation provides us with these useful relations which connect the spin coefficients along with the Weyl components. This result depends mainly on the form of the Killing Tensor since we demand by the rotation to preserve the Killing tensor invariant. Essentially, the annihilation of $\lambda_7$ allows us to obtain simplifications such that. 
 
 Thus, the key relations could help us to obtain the Petrov types of this solution initiating by the key relation (iv). The possible classes are three $\mu = 0\neq \tau$, $\mu = 0 =\tau$ and $\mu \neq 0 = \tau$.

\subsubsection{\it{Class I: $\mu = 0 \neq \tau$}}

The annihilation of $\mu$ has an immediate impact at key relations in the first place. Beginning with the condition $\tau \neq 0$, the first relation yields the annihilation of $\pi$ since $\Psi_2 =\Lambda$ due to relation (iii),

 \begin{equation}\tag{i} 0 = \pi \end{equation}
 \begin{equation}\tag{ii}\Psi_1 = 0 \end{equation}
 \begin{equation}\tag{iii}\Psi_2 - \Lambda = 0 \end{equation}
\begin{equation}\tag{iv}\mu = 0 \neq \tau\end{equation}

Afterwards, we will try to plug these into Newman-Penrose equations and into the integrability conditions. We take the annihilation of every Weyl component except $\Psi_0$. We already know that $\nu = 0$ nullifies $\Psi_4$. Next, the simultaneous nullification of $\pi$ and $\mu$ along with the NPE (m) and (h) clarifies that the solution is of type N since the only survivor is the $\Psi_0$.

\subsubsection{\it{Class II: $\mu = 0 = \tau$}}
The second class is determined by the following relations

\begin{equation}\tag{i} 0 = \tau \end{equation}
 \begin{equation}\tag{ii}\Psi_1 = 0 \end{equation}
 \begin{equation}\tag{iii}\Psi_2 - \Lambda = 0 \end{equation}
\begin{equation}\tag{iv}\mu = 0 = \tau\end{equation}
This combination does not make any difference. Again, we have nullification of $\Psi_4$ due to NPE (j). Additionally, the IC $(CR3:\lambda_1)$ along with NPE (i) nullifies $\Psi_3$ and the ($CR2:\lambda_2$) along with NPE (q) nullifies $\Psi_2$. This case is also characterized as Type N.

\subsubsection{\it{Class III: $\mu \neq 0 = \tau$}}
The key relations for the last class follow

\begin{equation}\tag{i} \Psi_2 - \Lambda = 0  \end{equation}
 \begin{equation}\tag{ii}\Psi_1 = \kappa \mu \end{equation}
 \begin{equation}\tag{iii} 0 = \rho \end{equation}
\begin{equation}\tag{iv}\mu \neq 0 = \tau\end{equation}
This case is a little trickier but it is proved to have the same Petrov type. The annihilation of $\nu$ gives $\Psi_4$. The IC $(CR4:\lambda_2)$ yields two possibilities,

\begin{equation}
    (\mu+\bar\mu)(\mu-\bar\mu)=0
\end{equation}
The only option is $\mu-\bar\mu =0$ since the simultaneous annihilation of $\rho$ and $\mu+\bar\mu$ gives $\lambda_2 =const$ resulting in a canonical form without two double eigenvalues. The latter along with NPE (k) annihilate $\Psi_1 $. Subsequently, the NPE (q) nullifies the $\Psi_2$ eventually, so the only non-zero Weyl components are $\Psi_0$ and $\Psi_3$. Initiating by NPE (i) and IC $(CR3:\lambda_1)$ we get

\begin{equation}\label{91}
    \Psi_3 = 2\pi \mu.
\end{equation}
By $(CR1:\lambda_2)$ and $(CR3:\lambda_2)$ we take  
\begin{equation}\label{92}
    \pi = \alpha +\bar\beta
\end{equation}
In this point we introduce the Bianchi identity (II) which yields with two  possible outcomes, 

\begin{equation}
    \kappa \Psi_3 = 0
\end{equation}
It is obvious that the annihilation of $\Psi_3$ gives the bespoken result. The Petrov type is determined as type N. But the other option, where $\kappa= Q(\bar\alpha+\beta +\bar\pi)=0$ along with (\ref{91}) and (\ref{92}) gives $\pi = \alpha +\bar\beta = 0$. Then, the space is characterized as conformally flat since the last relation annihilates $\Psi_3$ and $\kappa = 0 $ implying that $\Psi_0=0$ (NPE (\ref{b})).

\begin{table}[h]
		\centering 
			\caption{ Rotation around $\mu\tau=0$  }
		\begin{tabular}{ c c c }
			\toprule
			Type N & Type N & Type N   \\ 
			\toprule
   $\mu =0\neq \tau$  &  $\mu =0 =\tau$ &  $\mu \neq 0 =\tau $  \\ 
  $\Psi_0 \neq 0$     &$\Psi_0 \neq 0$     &$\Psi_0 \neq 0$       \\ 
 
  		\midrule
 $d\lambda_2 \neq 0$        & $d\lambda_2 \neq 0$    & $d\lambda_2 \neq 0$   \\ 
  	
			\bottomrule
		\end{tabular}
\end{table}
At last, the implication of the rotation provides us with simplifications paving the way to the following theorem.
\vspace{0.2cm}

\begin{theorem}
Assuming $K^1_{\mu \nu}$ canonical form of Killing tensor with $\lambda_7 = 0$ in vacuum with $\Lambda$ and applying a null rotation around the null tetrad frame Petrov type N solutions with a shear-free (non-diverging in one case) null congruence emerged.
\end{theorem}

\vspace{0.2cm}

 \section{ Solutions of $K^2_{\mu \nu} $and $K^3_{\mu \nu}$ form}\label{section 8}

Our objective is to determine the geometrical characterization (Petrov types) of the gravitational fields associated with the derived solutions, assuming the existence of $K^2$ and $K^3$ Killing forms with $\lambda_7=0$. In the next chapter we will exhibit the case where $\lambda_0 = 0$.  

\begin{equation} K_{(\mu \nu ; \alpha)}= 0 \end{equation}
Defining the factor $q= \pm 1$, we consider a unified approach for both canonical forms $K^2$ and $K^3$. The only difference in the $K^2$ and $K^3$ forms is the $-1$ in the $K_{22}$ component. Obviously, we get the $K^2_{\mu \nu}$ for $q=+1$ and the $K^3_{\mu \nu}$ for $q=-1$.

\begin{equation}K^{2,3}_{\mu \nu} = \lambda_0(n_\mu n_\nu + q l_\mu l_\nu)+ \lambda_1(l_\mu n_\nu + n_\mu l_\nu )     - \lambda_2 ( m_\mu \bar{m}_\nu +\bar{m}_\mu m_\nu)\end{equation}
This modification allows us to study the two forms simultaneously. The Killing equation of the Killing form yields the annihilation of $\sigma$ and $\lambda$ and the directional derivatives of $\lambda_0$, $\lambda_1$, $\lambda_2$.

\begin{equation}\sigma = \lambda = 0\end{equation}
\begin{equation}D \lambda_0 = 2 \lambda_0 (\epsilon + \bar{\epsilon}) \end{equation}
\begin{equation}\Delta \lambda_0 =   -2\lambda_0 (\gamma + \bar{\gamma}) \end{equation}  
\begin{equation}\label{defQ1}\delta \lambda_0 =  2\left[ \lambda_0 (\bar{\alpha} + \beta + \bar{\pi} ) - \kappa (\lambda_1 + \lambda_2) \right] \end{equation}
\begin{equation}\label{defQ2}\delta \lambda_0 =  2 \left[ -\lambda_0 (\bar{\alpha} + \beta + \tau ) +q\bar{\nu} (\lambda_1 + \lambda_2) \right] \end{equation}
\begin{equation}\delta \lambda_0 =  \lambda_0 (\bar{\pi} -\tau) -(\kappa -q\bar{\nu} )(\lambda_1 +\lambda_2)  \end{equation}

\begin{equation}D\lambda_1 = 2 \lambda_0 (\gamma + \bar{\gamma}) \end{equation}
\begin{equation}\Delta \lambda_1 = -2 q \lambda_0 (\epsilon + \bar{\epsilon}) \end{equation}
\begin{equation}\delta \lambda_1 =-q\lambda_0 (\kappa - q \bar{\nu}) +(\lambda_1 +\lambda_2)(\bar{\pi} - \tau)\end{equation}

\begin{equation}\label{Dl2} D\lambda_2 = \lambda_0 (\mu + \bar{\mu}) -(\lambda_1 +\lambda_2) (\rho +\bar{\rho}) \end{equation}
\begin{equation}\label{deltal2} \Delta \lambda_2 = -q\lambda_0 (\rho +\bar{\rho})   -(\lambda_1 +\lambda_2)   (\mu + \bar{\mu})\end{equation}
\begin{equation}\label{dl2}  \delta \lambda_2 = 0\end{equation}

Relations (\ref{defQ1}) and (\ref{defQ2}) allow us to define the factor $Q$.

\begin{equation}Q \equiv \frac{\lambda_0}{\lambda_1 +\lambda_2} = \frac {\kappa + q \bar{\nu}}{2 (\bar{\alpha} + \beta ) + \bar{\pi} + \tau } \end{equation}

\begin{equation}DQ = Q (2(\epsilon + \bar{\epsilon}) + (\rho + \bar{\rho})) -Q^2 ( 2(\gamma + \bar{\gamma}) + (\mu + \bar{\mu}) )\end{equation}
\begin{equation}\Delta Q = -Q (2(\gamma + \bar{\gamma}) + (\mu + \bar{\mu}))    +q Q^2  (2(\epsilon + \bar{\epsilon}) + (\rho + \bar{\rho}))\end{equation}
\begin{equation}\delta Q = (q Q^2 - 1 )(\kappa - q\bar{\nu})\end{equation}

The factor $Q$ is proved helpful during the treatment of the IC and it is a real scalar function since it depends solely on real scalars.

\vspace{0.2cm}

\subsection{Integrabillity Conditions of $K^2_{\mu \nu}$ and $K^3_{\mu \nu}$ forms}

We use the commutators of the tetrads in order to obtain the integrability conditions of Killing tensor. As we mentioned in subsection \ref{3.1}, the commutation relations are equivalent with the Lie bracket of the null tetrads. We choose to separate the integrability conditions using the factor $Q$. 

\vspace{0.2cm}

Integrability Conditions of $\lambda_0$
\footnotesize
\begin{equation}\tag{$CR1:\lambda_0$} 2Q [D (\gamma + \bar{\gamma}) + \Delta(\epsilon+\bar{\epsilon}) + \pi \bar{\pi} - \tau \bar{\tau} ] = - [(\pi + \bar{\tau}) (q\bar{\nu} - \kappa) +  (\bar{\pi}+\tau) ( q\nu - \bar{\kappa})]\end{equation} 
\begin{multline}\tag{$CR2:\lambda_0$} Q[ 2[\delta(\epsilon + \bar{\epsilon}) - (\epsilon+\bar{\epsilon}) (\bar{\alpha} +\beta - \bar{\pi})] -[D(\bar{\pi} -\tau) -(\bar{\pi} - \tau)(\bar{\rho} +\epsilon -\bar{\epsilon})]+2\kappa(\gamma +\bar{\gamma}) -(q\bar{\nu} - \kappa)[2(\gamma+\bar{\gamma})\\
 + (\mu +\bar{\mu})] ] = D(q\bar{\nu} - \kappa) - (q\bar{\nu} - \kappa) [2\epsilon +\bar{\rho}+ \epsilon +\bar{\epsilon} +\rho+\bar{\rho}] \end{multline}
\begin{multline}\tag{$CR3:\lambda_0$}Q[2[\delta(\gamma +\bar{\gamma}) +(\gamma +\bar{\gamma})(\bar{\alpha} +\beta - \tau)] +[\Delta(\bar{\pi} -\tau) +(\bar{\pi} - \tau)(\mu - \gamma+\bar{\gamma})] -2\bar{\nu}(\epsilon +\bar{\epsilon}) -q(q\bar{\nu} - \kappa)[2(\epsilon+\bar{\epsilon})\\
 +\rho+\bar{\rho}]] = \Delta(\kappa - q\bar{\nu}) +(\kappa - q\bar{\nu})[2(\gamma +\bar{\gamma}) +(\mu +\bar{\mu}) + \mu - \gamma+\bar{\gamma}] \end{multline}
\begin{multline}\tag{$CR4:\lambda_0$} Q[\bar{\delta}(\bar{\pi}-\tau) - \delta(\pi - \bar{\tau}) - (\bar{\pi} - \tau)(\alpha - \bar{\beta}) +(\pi - \bar{\tau})(\bar{\alpha} - \beta)+2[(\epsilon + \bar{\epsilon})(\mu-\bar{\mu}) -(\gamma +\bar{\gamma})(\rho-\bar{\rho})] ]\\
 = \delta(q\nu-\bar{\kappa}) - \bar{\delta}(q\bar{\nu} - \kappa) +(q\bar{\nu}-\kappa)(\alpha-\bar{\beta}) - (q\nu - \bar{\kappa})(\bar{\alpha}-\beta)\end{multline}
\vspace{0.2cm}
\normalsize

Integrability Conditions of $\lambda_1$ 

\footnotesize
\begin{equation}\tag{$CR1:\lambda_1$}Q[ \Delta(\gamma + \bar{\gamma}) - 3(\gamma+\bar{\gamma})^2 +q[ D(\epsilon+\bar{\epsilon}) +3(\epsilon+\bar{\epsilon})^2] +\frac{q}{2}[ (\pi + \bar{\tau}) (q\bar{\nu} - \kappa) +  (\bar{\pi}+\tau) ( q\nu - \bar{\kappa})]] 
=-(\pi \bar{\pi} - \tau\bar{\tau}) 
\end{equation}
\begin{multline}\tag{$CR2:\lambda_1$} Q[ 2[ \delta(\gamma+\bar{\gamma}) -( \gamma + \bar{\gamma})(\bar{\alpha} + \beta-\bar{\pi}) ]  -q[D(q\bar{\nu}-\kappa) +(q\bar{\nu}-\kappa)(\epsilon + 3\bar{\epsilon} + \bar{\rho}) - 2\kappa(\epsilon+\bar{\epsilon})]  ] \\
= D(\bar{\pi}-\tau) - (\bar{\pi} - \tau)(\rho + 2\bar{\rho} +\epsilon-\bar{\epsilon}) - 2(\gamma + \bar{\gamma})(q\bar{\nu} - \kappa)
\end{multline}
\begin{multline}\tag{$CR3:\lambda_1$}Q[2q[\delta(\epsilon+\bar{\epsilon}) + (\epsilon +\bar{\epsilon})(\bar{\alpha}+ \beta - \tau)]+q[\Delta(q\bar{\nu}-\kappa) - (q\bar{\nu} - \kappa)(3\gamma+\bar{\gamma} -\mu) ] -2\bar{\nu}(\gamma+\bar{\gamma}) ]\\
= -[\Delta(\bar{\pi} - \tau) + (\bar{\pi}-\tau)(2\mu +\bar{\mu}-\gamma+\bar{\gamma})+2q(q\bar{\nu} - \kappa)(\epsilon+\bar{\epsilon})]
\end{multline}
\begin{multline}\tag{$CR4:\lambda_1$} Q[q[\delta(q\nu-\bar{\kappa}) - \bar{\delta}(q\bar{\nu} - \kappa) +(q\bar{\nu}-\kappa)(\alpha-\bar{\beta}) - (q\nu - \bar{\kappa})(\bar{\alpha}-\beta)] +2[q(\epsilon+\bar{\epsilon})(\rho - \bar{\rho}) - (\gamma+\bar{\gamma})(\mu-\bar{\mu})] \\
=\bar{\delta}(\bar{\pi}-\tau) - \delta(\pi - \bar{\tau}) - (\bar{\pi} - \tau)(\alpha - \bar{\beta}) +(\pi - \bar{\tau})(\bar{\alpha} - \beta) \end{multline}
\vspace{0.2cm}
\normalsize

Integrability Conditions of $\lambda_2$ 

\footnotesize
\begin{multline}\tag{$CR1:\lambda_2$} Q[ [\Delta(\mu+\bar{\mu})-(\mu+\bar{\mu})-5(\gamma+\bar{\gamma})]+q[D(\rho+\bar{\rho}) + (\rho+\bar{\rho})[(\rho+\bar{\rho})-5(\epsilon+\bar{\epsilon})]] ] \\
= \Delta(\rho+\bar{\rho}) -(\rho+\bar{\rho})(\gamma+\bar{\gamma}) + D(\mu+\bar{\mu}) +(\mu+\bar{\mu})(\epsilon+\bar{\epsilon}) \end{multline}
\begin{multline}\tag{$CR2:\lambda_2$}Q[ \delta(\mu+\bar{\mu}) -(\mu+\bar{\mu})[(\bar{\alpha}+\beta+\tau)-2\bar{\pi}] +q(\rho+\bar{\rho})(2\kappa - q\bar{\nu}) ]\\
=\delta(\rho+\bar{\rho}) -(\rho+\bar{\rho})[\bar{\alpha}+\beta+\tau-2\bar{\pi}] + (\mu+\bar{\mu})(2\kappa-q\bar{\nu})\end{multline}
\begin{multline}\tag{$CR3:\lambda_2$} qQ[\delta(\rho+\bar{\rho})+(\rho+\bar{\rho})[\bar{\alpha}+\beta+\bar{\pi}-2\tau] +(\mu+\bar{\mu})(\kappa-2q\bar{\nu})] \\
=\delta(\mu+\bar{\mu})+(\mu+\bar{\mu})[\bar{\alpha} +\beta+\bar{\pi}-2\tau]+q(\rho+\bar{\rho})(\kappa -2q\bar{\nu}) \end{multline}
\begin{multline}\tag{$CR4:\lambda_2$} Q[(\mu+\bar{\mu})(\mu-\bar{\mu}) -q(\rho+\bar{\rho})(\rho-\bar{\rho})] =   (\mu-\bar{\mu})(\rho+\bar{\rho}) - (\rho-\bar{\rho})(\mu+\bar{\mu})
\end{multline}
\normalsize

\subsection{Classes of solutions}
We choose to take advantage of the conformal symmetry of a rotation around one of the real null vectors (Chapter \ref{section 4}). Subsequently, we require that the Killing tensor remains invariant under rotation, as discussed. 
$$K = \lambda_0 (\tilde{\theta}^1 \otimes \tilde{\theta}^1 +q \tilde{\theta}^2 \otimes \tilde{\theta}^2 ) +\lambda_1(\tilde{\theta}^1 \otimes \tilde{\theta}^2+\tilde{\theta}^2 \otimes \tilde{\theta}^1) + \lambda_2(\tilde{\theta}^3 \otimes \tilde{\theta}^4+\tilde{\theta}^4 \otimes \tilde{\theta}^3); \hspace{0.5cm} q= \pm 1$$
It is easy for someone to prove that the only non-zero rotation parameter is $t=ib$. As a matter of fact, we obtain the exact same key relations since the null rotation is applied with the exact same way as in the previous chapter. 

Moving forward, we shall present the classes of solutions are classified by the usage of key relations. Initiating by  $\mu \tau = 0 $ the NPE, IC, BI, provide us with three main classes of solutions. Also, we aim to study solutions in which only one of the components of Killing tensor, is allowed to be constant. It would be helpful for the reader if at this point we array the most useful NPE and the $(CR4:\lambda_2)$. 

\begin{equation}\tag{b} \delta \kappa = \kappa ( \tau - \bar{\pi} +2(\bar{\alpha} +\beta) - (\bar{\alpha} - \beta) ) - \Psi _o  \end{equation}
\begin{equation}\tag{g} \bar{\delta} \pi = - \pi(\pi + \alpha - \bar{\beta}) + \nu\bar{\kappa} \end{equation}
\begin{equation}\tag{p} \delta \tau = \tau (\tau - \bar{\alpha} + \beta) -\bar{\nu} \kappa  \end{equation}
\begin{equation}\tag{k} \delta\rho = \rho(\bar{\alpha} + \beta) +\tau(\rho-\bar{\rho})+ \kappa(\mu-\bar{\mu}) - \Psi_1 \end{equation}
\begin{equation}\tag{m} \bar{\delta} \mu = -\mu (\alpha +\bar{\beta}) -\pi (\mu- \bar{\mu}) - \nu (\rho-\bar{\rho}) + \Psi_3 \end{equation}
\begin{equation}\tag{j} \bar{\delta}\nu = - \nu(2(\alpha +\bar{\beta} ) + (\alpha - \bar{\beta})  + \pi - \bar{\tau}  ) + \Psi_4\end{equation}
\small
\begin{multline}\tag{$CR4:\lambda_2$} Q[(\mu+\bar{\mu})(\mu-\bar{\mu}) -q(\rho+\bar{\rho})(\rho-\bar{\rho})] = (\mu-\bar{\mu})(\rho+\bar{\rho})-(\rho-\bar{\rho})(\mu+\bar{\mu})
\end{multline}
\normalsize

 \vspace{0.1cm}

 \subsubsection{\it{Class I : $\mu = 0$}}
 
  The annihilation of $\mu$ gives the first class of solution. Considering that $\mu=0$ the relation $(CR4:\lambda_2)$ gives $(\rho +\bar{\rho})(\rho-\bar{\rho})=0$. According to IC (\ref{Dl2})-(\ref{dl2}) the derivative of the eigenvalue $\lambda_2$ depends only from the real parts of $\mu$ and $\rho$, so our priority is to avoid $\rho +\bar{\rho} = 0$. Unavoidably, the BI (\ref{II}) implies that the derivative of $\lambda_2$ vanishes. However, from NPE (\ref{m}) we take $\Psi_3 = \nu (\rho - \bar{\rho}) = 0 $. 

In the following Tables \ref{table1}, \ref{table2} every column represents different solutions according to different choices which are ordered by the key relations and the BI (II). The columns of the tables contain the main characteristics of our solutions. The second and third column of Table \ref{table1} is distinguished by the different choices that take place due to the BI (III) and BI (VI).
 \begin{table}[h] 
			\caption{ $\rho -\bar{\rho} =  0 = \rho+\bar{\rho}$}\label{table1}
		\begin{tabular}{ c c c }
			\toprule
			Type D& Type N& Type N  \\			
   $\rho= 0 \neq \Psi_2$ & $\rho + \bar{\rho} = 0 = \Psi_2$ & $\rho + \bar{\rho} = 0 = \Psi_2$  \\
			\toprule
$ \kappa \nu = \pi \tau$   & $ \nu =0 =  \pi \tau  $ & $  \kappa =0 =  \pi \tau$ \\
$\Psi_0 \Psi_4 = {9\Psi_2}^2$ & $\Psi_0\neq0 $    &$\Psi_4\neq0 $\\
 $\Psi_1= \Psi_3 = 0$  &$ \Psi_1= \Psi_2 = \Psi_3 = \Psi_4 = 0$  & $\Psi_0 = \Psi_1= \Psi_2 =\Psi_3 = 0$ \\
			\midrule
 $d\lambda_2 = 0$     & $d\lambda_2 = 0$ &    $d\lambda_2 = 0$ \\ 
			\bottomrule
		\end{tabular}
	
 \end{table}
The other choice where $\rho+\bar{\rho}\neq0=\Psi_2$ yields type N solutions only where the eigenvalue $\lambda_2$ is not a constant. The combination of BI (III) with BI (VI) determines the non-zero Weyl component.
\begin{table}[h]
 
        \caption{ $\rho -\bar{\rho} =  0 \neq \rho+\bar{\rho}$ }\label{table2}
		\begin{tabular}{ c c c c c c }
			\toprule
			Type N & Type N & Type N &Type N &Type N & Type N   \\ 
			\toprule
   $\nu =0= \tau$  &  $\nu =0 =\pi$ &  $\kappa =0 =\tau $ & $  \kappa =0 =\pi $ &  $ \nu = \pi= \tau=0$ & $ \kappa = \pi =\tau = 0$  \\ 
  $\Psi_0\neq 0$            & $\Psi_4 \neq 0$& $\Psi_0\neq 0$            & $\Psi_4 \neq 0$&  $\Psi_0\neq 0$     & $\Psi_4 \neq 0$  \\ 
 
  		\midrule
 $d\lambda_2 \neq 0$     &  $d\lambda_2 \neq 0$  & $d\lambda_2 \neq 0$     & $d\lambda_2 \neq 0$     & $d\lambda_2 \neq 0$    & $d\lambda_2 \neq 0$   \\ 
  		
			\bottomrule
\end{tabular}

\end{table}

\subsubsection{\it{Class II: $\mu = 0 = \tau$}}
 As before, the NPE (\ref{m}) yields $\Psi_3 = \nu (\rho - \bar{\rho}) = 0$. Considering the key relation (ii) we obtain $\Psi_1 = 0$, similarly considering NPE (\ref{i}) we obtain $\kappa \nu = 0$. Thus, it is obvious that the Class II consists a subset of Class I. These solutions are the same type N solutions of the previous class with the only difference to be that $\tau = 0$.
 
\subsubsection{\it{Class III: $\tau = 0$}}
In this class we encountered new algebraically special solutions. NPE (\ref{p}) for $\tau =0$ yields $\bar{\kappa} \nu = 0$. The branch where $\mu = 0$ is already known from the other two classes where concerns type N solutions with $\tau=0$. On the other hand, the case of $\mu \neq  0 = \rho $ yields two Type III solutions. 

The IC $(CR4:\lambda_2)$ plays a crucial role in this scene since the annihilation of $\rho$ implies $(\mu +\bar{\mu})(\mu - \bar{\mu}) = 0$. Although, the constraint $\mu \neq 0$ leads us to two separate solutions for case $\kappa = 0 \neq \nu$, which are both of Type III. The other case, where both $\kappa$ and $\nu$ are zero, concerns two solutions where only $\Psi_3$ is not equal to zero determining that the type of the solutions are also of Type III.

The last branch of solutions contains the case $\kappa \neq 0 = \nu$. It should be noted that NPE (\ref{k}) with $\rho = 0$ determines $\Psi_1$, which annihilates the real part of $\mu$ which is equivalent with $d\lambda_2 = 0$. Then NPE (k) is given by the following relation.

\begin{equation}\tag{k} \Psi_1 = \kappa (\mu - \bar{\mu}) \end{equation}

\begin{table}[h]
			\caption{ $\mu \neq  0 = \rho$ }\label{table3}
		\begin{tabular}{ c c c c }
			\toprule
			$\mu - \bar{\mu}= 0 $ & $\mu + \bar{\mu} = 0$ &$\mu - \bar{\mu}= 0 $ & $\mu + \bar{\mu} = 0$  \\
			Type III& Type III &Type III& Type III \\
			\toprule
$ \kappa=0\neq \nu $   & $ \kappa =0 \neq \nu  $ &$ \kappa=0= \nu $   & $ \kappa =0 = \nu  $ \\
$\Psi_3\neq 0 \neq \Psi_4 $ & $\Psi_3\neq0 \neq \Psi_4$& $\Psi_3 \neq 0 = \Psi_4$ &$\Psi_3\neq 0 = \Psi_4 $   \\
			\midrule
 $d\lambda_2 \neq 0$     &  $d\lambda_2 = 0$&$d\lambda_2 \neq 0$     &  $d\lambda_2 = 0$  \\ 
			\bottomrule
		\end{tabular}
	\end{table}

The Class III is presented in Table \ref{table3}. All these cases are characterized by $\Psi_2 = \Lambda=0$ which arises from NPE (q) for $\rho = \tau = 0$, while in these type III solutions $\Psi_1 = 0$. Finally, we postulate the following two theorems. 
\vspace{0.4cm}

\begin{theorem}
Assuming $K^2_{\mu \nu}, K^3_{\mu \nu}$ canonical forms of Killing tensor with $\lambda_7 = 0$ in vaccum with $\Lambda$ and applying a null rotation around the null tetrad frame Petrov types III, N solutions with a shear-free, diverging (and in some cases geodesic) null congruence emerged.
\end{theorem}

\begin{theorem}\label{theorem 4}

Assuming $K^2_{\mu \nu}, K^3_{\mu \nu}$ canonical forms of Killing tensor with $\lambda_7 = 0$ in vacuum with $\Lambda$ and applying a null rotation around the null tetrad frame a unique Petrov type D solution with a shear-free, diverging and non-geodesic null congruence emerged.
\end{theorem}

\section{Solutions of $K^4_{\mu \nu}$ form}\label{section 9}

In line with the previous chapters we study one more form which does not belong to the four canonical forms of Killing tensor although is a subcase of $K^1_{\mu \nu}, K^2_{\mu \nu}, K^3_{\mu \nu}$ forms and it is obtained when $\lambda_0$ equals to zero. 
We wanted to investigate the virtue of the solutions that one could obtain using the other possible reduction of Killing forms $K^1, K^2, K^3$, namely, when $\lambda_0=0$.

With reference to Remark \ref{remark} we are able to obtain somehow different simplifications due to the rotation around the null tetrad frame. We initiate our analysis defining the factor $Q$ in the same fashion as in Chapter \ref{section 6}.

\begin{equation}
    Q \equiv \frac{\lambda_7}{\lambda_1 + \lambda_2}
\end{equation}
The Killing form is represented in terms of the null tetrads as follows.   
\begin{equation}
    K^4_{\mu \nu} = \lambda_1(n_\mu l_\nu + l_\mu n_\nu) + \lambda_2(m_\mu \bar{m}_\nu + \bar{m}_\mu m_\nu) +\lambda_7(m_\mu m_\nu +\bar{m}_\mu \bar{m}_\nu)
\end{equation}
The Killing equation yields

\begin{equation}\label{129}
    \kappa(\lambda_1 +\lambda_2) +\bar\kappa\lambda_7 =0
\end{equation}
\begin{equation}\label{130}
    \nu(\lambda_1 +\lambda_2) +\bar\nu\lambda_7 =0
\end{equation}
\begin{equation}
    D\lambda_1 = \Delta \lambda_1 =0 
\end{equation}
\begin{equation}
    \delta\lambda_1 =(\bar\pi -\tau)(\lambda_1+\lambda_2) +(\pi -\bar\tau)\lambda_7 
\end{equation}
\begin{equation}
    D\lambda_2= -(\rho+\bar\rho)(\lambda_1+\lambda_2 -(\sigma +\bar\sigma)\lambda_7
\end{equation}
\begin{equation}
    \Delta\lambda_2 = (\mu +\bar\mu)(\lambda_1+\lambda_2) +(\lambda+\bar\lambda)\lambda_7
\end{equation}
\begin{equation}
    \delta \lambda_2 = 2\lambda_7(\alpha -\bar\beta)
\end{equation}
\begin{equation}
    D\lambda_7 = -2\bar\sigma(\lambda_1+\lambda_2) -2\rho\lambda_7
\end{equation}
\begin{equation}
    \Delta \lambda_7 = 2\lambda(\lambda_1+\lambda_2) +2\bar\mu\lambda_7
\end{equation}
\begin{equation}
    \delta\lambda_7 =-2\lambda_7(\bar\alpha - \beta) 
\end{equation}

\subsection{Integrability Conditions of $K^4_{\mu \nu }$ form}

Integrability Conditions of $\lambda_1$
\footnotesize
\begin{equation}\tag{$CR1:\lambda_1$} 2(\pi\bar\pi - \tau\bar\tau) + Q\left[ (\pi+\bar\tau)(\pi -\bar\tau)+(\bar\pi+\tau)(\bar\pi-\tau) \right] = 0 \end{equation} 
\begin{multline}\tag{$CR2:\lambda_1$} D(\bar\pi-\tau) - (\bar\pi-\tau)(\rho+2\bar\rho)-(\pi -\bar\tau)(\sigma +2\bar\sigma) 
\\+Q \left[ D(\pi-\bar\tau) - (\bar\pi-\tau)(2\sigma+\bar\sigma) -(\pi-\bar\tau)(2\rho+\bar\rho) \right] = 0   \end{multline}
\begin{multline}\tag{$CR3:\lambda_1$} \Delta(\bar\pi-\tau) +(\bar\pi-\tau)(2\mu+\bar\mu)+(\pi-\bar\tau)(2\lambda+\bar\lambda)
\\+Q\left[ \Delta(\pi-\bar\tau) +(\bar\pi-\tau)(\lambda+2\bar\lambda)+(\pi-\bar\tau)(\mu +2\bar\mu)\right] = 0 \end{multline}
\begin{equation}\tag{$CR4:\lambda_1$} \bar\delta(\bar\pi-\tau) - \delta(\pi-\bar\tau) +Q\left[\bar\delta(\pi-\bar\tau) - \delta(\bar\pi-\tau) +(\bar\pi-\tau)^2 -(\pi-\bar\tau)^2 \right] = 0 \end{equation}
\vspace{0.2cm}
\normalsize

Integrability Conditions of $\lambda_2$ 

\footnotesize
\begin{multline}\tag{$CR1:\lambda_2$} \Delta(\rho+\bar\rho) +D(\mu +\bar\mu) -(\gamma+\bar\gamma)(\rho+\bar\rho) +(\epsilon+\bar\epsilon)(\mu+\bar\mu)+2(\lambda\sigma - \bar\lambda\bar\sigma) \\
+Q\left[\Delta(\sigma+\bar\sigma)+ D(\lambda +\bar\lambda) +(\lambda+\bar\lambda)(\epsilon+\bar\epsilon -\rho+\bar\rho) -(\sigma+\bar\sigma)(\gamma+\bar\gamma+\mu-\bar\mu) \right] = 0
\end{multline}
\begin{multline}\tag{$CR2:\lambda_2$} \delta(\rho+\bar\rho)+(\rho+\bar\rho)(2\bar\pi-\tau -\bar\alpha-\beta) +\kappa(\mu +\bar\mu) \\ 
+Q\left[ \delta(\sigma+\bar\sigma) +(\pi-\bar\tau)(\rho+\bar\rho) -(\sigma+\bar\sigma)(\bar\alpha+\beta - \bar\pi) +\kappa(\lambda+\bar\lambda) \right]  = 0 
\end{multline}
\begin{multline}\tag{$CR3:\lambda_2$} \delta(\mu +\bar\mu) +(\mu +\bar\mu)(\bar\pi-2\tau+\bar\alpha+\beta) -\bar\nu(\rho+\bar\rho)\\
Q\left[\delta(\lambda+\bar\lambda) +(\mu+\bar\mu)(\bar\pi-\tau) -(\lambda+\bar\lambda)(\tau- \bar\alpha-\beta) - \bar\nu(\sigma+\bar\sigma) \right] = 0
\end{multline}
\begin{equation}\tag{$CR4:\lambda_2$}(\mu-\bar\mu)(\rho+\bar\rho)-(\rho-\bar\rho)(\mu+\bar\mu) +Q\left[(\mu-\bar\mu)(\sigma+\bar\sigma)-(\rho-\bar\rho)(\lambda+\bar\lambda)\right] = 0\end{equation}
\vspace{0.2cm}
\normalsize

Integrability Conditions of $\lambda_7$ 

\footnotesize
\begin{equation}\tag{$CR1:\lambda_7$} \Delta \bar\sigma +D\lambda -\bar\sigma(\gamma+\bar\gamma-\mu+\bar\mu) +\lambda(\epsilon+\bar\epsilon +\rho-\bar\rho) +Q\left[D\bar\mu +\Delta\rho - (\lambda\sigma-\bar\lambda\bar\sigma) +\bar\mu(\epsilon+\bar\epsilon) -\rho(\gamma+\bar\gamma) \right]=0 \end{equation}
\begin{equation}\tag{$CR2:\lambda_7$}\delta\bar\sigma +\bar\sigma(2\bar\pi-\tau -\bar\alpha-\beta) +\kappa\lambda +Q\left[\delta\rho+\bar\sigma(\pi-\bar\tau)-\rho(\bar\alpha+\beta-\bar\pi) +\kappa\bar\mu \right] =0  \end{equation}
\begin{equation}\tag{$CR3:\lambda_7$} \delta\lambda +\lambda(\bar\pi-2\tau+\bar\alpha+\beta)- \bar\sigma\bar\nu +Q\left[ \delta\bar\mu +\lambda(\pi-\bar\tau)-\rho\bar\nu -\bar\mu(\tau-\bar\alpha-\beta)\right] = 0\end{equation}
\begin{equation}\tag{$CR4:\lambda_7$} \bar\sigma(\mu-\bar\mu) - \lambda(\rho-\bar\rho) +Q\left[ \rho(\mu-\bar\mu)-\bar\mu(\rho-\bar\rho)\right]= 0 \end{equation}
\normalsize

\subsection{Classes of solutions}
Similarly, we choose $l^\mu$ to be fixed during the rotation. Afterward, we demand that the Killing tensor remains invariant under the rotation but in this case the non-zero rotation parameter is proved to be $t=a$ instead of $t=ib$. The latter provides us with different simplifications due to the annihilation of the tilded spin coefficients. Namely, we get the \textit{key relations} which unfold the branches of the solutions,    

 \begin{equation}\tag{i} \Psi_2 - \Lambda = \kappa \nu - \tau \pi \end{equation}
 \begin{equation}\tag{ii}\Psi_1 = \kappa \mu -\sigma\pi\end{equation}
 \begin{equation}\tag{iii}\Psi_2 - \Lambda = \mu \rho - \sigma\lambda\end{equation}
\begin{equation}\tag{iv}\mu \tau -\sigma\nu= 0\end{equation}

Also, considering the last four tilded spin coefficients we get:

\begin{equation}
    \epsilon - \bar\epsilon = 0
\end{equation}
\begin{equation}
\gamma- \bar\gamma=0    
\end{equation}
\begin{equation}
    \alpha - \bar\beta = 0
\end{equation}

The last three relations marks the difference compared to the simplifications made in the previous chapters, due to the exact same rotation. Based on the \textit{key relation} of $\Psi_1 = \kappa\mu-\sigma\pi$ and the relation $\alpha-\bar\beta=0$ we can easily derive the two following relations. The first one was obtained by the subtraction of the complex conjugate of NPE (d) with NPE (e) and the second one by the summation of the complex conjugate of NPE (r) with NPE (o).

\begin{equation}\label{d,e} \pi\rho -\kappa\lambda = 0\end{equation}
\begin{equation}\label{r,o} \Psi_3 = \nu\rho - \lambda(\tau+\beta)
\end{equation}

By consideration of relations (\ref{129}), (\ref{130}) we choose to annihilate the spin coefficients $\kappa, \nu$ avoiding to correlate the elements of Killing tensor between themselves obtaining either $\lambda_1 + \lambda_2 +\lambda_7 = 0 $ or $\lambda_1 +\lambda_2 - \lambda_7 = 0$. We made this choice since our goal for this chapter is to study a Killing tensor with one double eigenvalue. Moreover, the $K^4_{\mu \nu}$ form has 3 eigenvalues where $\lambda_1$ to be double and the others to be equal to $-(\lambda_2 + \lambda_7)$ and $-(\lambda_2-\lambda_7)$\footnote{However, we briefly examined the case where $Q+1=0$ for $K^4_{\mu \nu}$ form proving that this special case admit Petrov types III, N, O in contrast with $K^0_{\mu \nu}$ of Chapter 5. The proof of the latter is not given in this paper.}. The simplifications of spin coefficients till now are presented.

\begin{equation}
    \kappa = \nu = \epsilon-\bar\epsilon =\gamma-\bar\gamma = \alpha-\bar\beta = 0
\end{equation}

Regarding this, three classes of solution emerge by relation (\ref{d,e}). Either $\pi = 0$ or $\rho = 0$ or both. 

\subsubsection{\textit{Class I: $ \rho \neq 0 =\pi$}}
 This first class of solution results in three solutions, two of type N solutions and one of type D solution. The relation $(\ref{r,o})$ and NPE (i) give $\alpha =\bar\beta = 0$. Also, the key relations reads  

 \begin{equation}\tag{i} \Psi_2 - \Lambda = 0 \end{equation}
 \begin{equation}\tag{ii}\Psi_1 = 0 \end{equation}
 \begin{equation}\tag{iii}\Psi_2 - \Lambda = 0= \mu \rho - \sigma\lambda\end{equation}
\begin{equation}\tag{iv}\mu \tau = 0\end{equation}

where the last of them yields three possibilities. The \textbf{first case} of these possibilities is obtained when $\mu=0\neq\tau$.

\begin{equation}\tag{n} \lambda = 0 
\end{equation}
\begin{equation}\tag{h}
    \Psi_2 = \Lambda = 0
\end{equation}
\begin{equation}\tag{j}
    \Psi_4 = 0
\end{equation}

The relation $(\ref{r,o})$ and the key relation (ii) annihilate $\Psi_3$ and $\Psi_1$ accordingly. The results of the first case are summarized in the two following relations which characterize our solution as type N.

$$\kappa = \nu = \pi=\alpha = \beta = \mu = \lambda =\epsilon -\bar\epsilon = \gamma-\bar\gamma= 0$$
$$\Psi_1 = \Psi_2 = \Psi_3 = \Psi_4 = 0$$

The \textbf{second case} where $\mu = 0 = \tau$ is also of type N indicating that it is a subset of the first case. 

We shall proceed to the \textbf{third case} which is the most interesting one since it is proved to be type D when $\mu\neq0=\tau$. It is also important to show that the type D character of this solution emerged by combining the key relation (iii) $\mu\rho-\sigma\lambda = 0$ with the BI (II) and (VII).
\begin{equation}\tag{II}
   0=  3\Psi_2 \rho -\lambda \Psi_0
\end{equation}
\begin{equation}\tag{VII}
    0=\sigma\Psi_4 - 3\Psi_2\mu
\end{equation}

Hence, dividing by parts we obtain the already known combination of the Weyl components as in the previous chapter.
\begin{equation}
    \Psi_0\Psi_4 - (3\Psi_2)^2 = 0
\end{equation}

The following relations encapsulate the nullified spins of the third case.

$$\kappa = \nu = \pi = \tau=\alpha=\beta=\epsilon -\bar\epsilon = \gamma-\bar\gamma= 0$$
$$\Psi_1 = \Psi_3 = 0$$
$$\Psi_0\Psi_4 = (3\Psi_2)^2 = (3\Lambda)^2$$

The  Table 5 contains the three solutions of Class  I that we presented in this section. 

\begin{equation}
    \kappa = \nu = \pi= \alpha = \beta =\epsilon-\bar\epsilon=\gamma-\bar\gamma=0 
\end{equation}
It is neccessary to denote that for any of the following cases the previous spins annihilation hold.

\vspace{0.3cm}

\begin{table}[h] 
			\caption{ Class I: $\rho\neq0=\pi$}\label{table5}
		\begin{tabular}{ c c c }
   \toprule
			Type N& Type N& Type D  \\			
   $\mu= 0 \neq \tau$ & $\mu = 0 = \tau$ & $\mu\neq 0 = \tau$  \\
			\midrule
$ \mu=\lambda= 0$   & $\mu=\lambda=\tau= 0$ & $ \tau= 0$ \\
$\Psi_0 \neq0$ & $\Psi_0\neq0 $    &$\Psi_0\Psi_4-(3\Psi_2)^2=0 $\\

\bottomrule
\end{tabular}
	
 \end{table}

\subsubsection{\textit{Class II: $ \rho = 0 =\pi$}}

Initiating by $\kappa=\nu=\pi=\rho = 0$ the key relation reads

 \begin{equation}\tag{i} \Psi_2 - \Lambda = 0 \end{equation}
 \begin{equation}\tag{ii}\Psi_1 = 0\end{equation}
 \begin{equation}\tag{iii}\Psi_2 - \Lambda =0= \sigma\lambda\end{equation}
\begin{equation}\tag{iv}\mu \tau = 0\end{equation}

the only non-conformally flat solution arises when $\mu\neq0=\tau$. The NPE reads   

\begin{equation}\tag{a}
    \sigma=0
\end{equation}
\begin{equation}\tag{b}
    \Psi_0=0
\end{equation}
\begin{equation}\tag{c}
    \Psi_1=0
\end{equation}
\begin{equation}\tag{i}
    \Psi_3 = - \lambda\beta = 0
\end{equation}
\begin{equation}\tag{q}
    \Psi_2=\Lambda = 0
\end{equation}
\begin{equation}\tag{l}
    \epsilon(\mu-\bar\mu)=0
\end{equation}
\begin{equation}\tag{j}
    -\Delta\lambda = \lambda(\mu+\bar\mu+2\gamma) +\Psi_4
\end{equation}

A type N solution is achieved only when $\alpha=\beta=0$ according to the NPE (i). All other scenarios result in type O solution, which does not pique our interest.

\subsubsection{\textit{Class III: $ \rho = 0 \neq\pi$}}
The key equation for this class is slightly different.
 \begin{equation}\tag{i} \Psi_2 - \Lambda = -\pi\tau \end{equation}
 \begin{equation}\tag{ii}\Psi_1 = -\pi\sigma\end{equation}
 \begin{equation}\tag{iii}\Psi_2 - \Lambda =-\sigma\lambda \end{equation}
\begin{equation}\tag{iv}\mu \tau = 0\end{equation}

Implying the annihilation of $\sigma$ due to NPE (b) we get $\Psi_0=0$, also the key relation (i) reads

\begin{equation}\tag{i}
    \Psi_2-\Lambda = 0 = \pi\tau \hspace{0.2cm} \longrightarrow \hspace{0.1cm} \tau = 0 \hspace{0.1cm};\hspace{0.15cm} \pi\neq0
\end{equation}
Furthermore, implying the latter to NPE (k) and to NPE (q) we obtain $\Psi_1 =0$ and $\Psi_2=\Lambda=0$ accordingly. Thus, the only non-zero Weyl component is $\Psi_3=-\lambda\beta$ indicating that this solution is of Petrov type III.

\vspace{0.2cm}

\begin{theorem}\label{theorem 5}
   Assuming $K^4_{\mu \nu}$ canonical form of Killing tensor with $\lambda_0=0$ in vacuum with $\Lambda$ and applying a null rotation around the null tetrad frame Petrov type N and D solutions with a geodesic, shearing and diverging null congruence emerged. 
\end{theorem}

\begin{theorem}
    Assuming $K^4_{\mu \nu}$ canonical form of Killing tensor with $\lambda_0=0$ in vacuum with $\Lambda$ and applying a null rotation around the null tetrad frame Petrov type N and III solutions with a geodesic, non-shearing and non-diverging null congruence emerged.
\end{theorem}

\section{Discussion and Conclusions}\label{Section 10}

The \textit{Study of the Canonical forms of Killing tensor} is initiated by considering the possibility to obtain more interesting spacetimes with hidden symmetries using the canonical forms as an initial premise. In order to prove this statement properly we should rather have extracted the corresponding solutions that admit the canonical forms of Killing tensor in electro-vacuum, where the Einstein-Maxwell tensor $F_{\mu \nu}$ would be present and comparing to those of the work of Hauser-Malhiot \cite{Hauser1976}. This work will be conducted in the immediate future. However, quite useful conclusions were extracted during this investigation (Vacuum with $\Lambda$) concerning this hypothesis.

In an attempt to address the aforementioned hypothesis and provide a proper answer, we sought to establish a \textit{rule} to counterbalance the arbitrariness of the tetrad frame. In pursuit of solutions of the most general Petrov type, the authors believe that the transformation process must be as general as possible providing the most general Petrov type. As far as we know, the transformations applied do not follow any particular rule, and there is no any directive regarding the most general transformative process. For this reason, this study was operated scoping to imply only a null rotation around $l^\mu$,\footnote{The same key relations would be obtained if $n^\mu$ was fixed.} thus, the resolution process was attained under this \textit{rule}. Regarding this, only $p=c+id$ and $a$ (boost) or $ib$ (twist) were annihilated due to the conservation of Killing tensor. Afterwards, none simplification takes place leaving the remained rotation parameters undetermined. 

According to Remark \ref{remark}, the null rotation is proved fruitful only when the reduced forms of Killing tensor are present since the invariant character of the Killing tensor would annihilate both $t=a+ib$ and $p = c+id$ ``draining" the remaining tetrad freedom. Strictly speaking, the capitalization of the rotation parameters manages to gain the $\textbf{key relations}$ which basically reveal the classes of our solutions. Differently, we could take advantage of the free rotation parameter $t$ correlating the spin coefficients within themselves (Lorentz transformations) only. In other words, the only way to obtain simplifications is either to employ reduced Killing tensor or to imply Lorentz transformations correlating the spin coefficients between themselves which is equivalent to the determination of $t$.

The latter choice falls into the concept of the \textit{symmetric null tetrads} that was manifested in \cite{debever1981orthogonal}. Using the free rotation parameters $a$ and $b$ Debever managed to obtain the following simplifications that led him to the most general type D solution with a geodesic, shear-free null congruence \cite{debever1984exhaustive} assuming that the PND of Einstein-Maxwell tensor aligned to the PND of Weyl tensor. Namely, the following relations determine the rotation parameters $a$ and $b$

$$\pi=e\tau$$
$$\mu = e\rho$$
$$\gamma= e\epsilon$$
$$e=\pm1$$

The interesting part arises when we derived the Petrov types of the extracted solutions assuming the existence of the Killing tensor of our \textit{Paradigm} both ways. Firstly we implied the symmetric null tetrad concept, afterwards the null rotation methodology of Chapter \ref{section 4}. With fundamental derivations it is really easy to prove that using the symmetric null tetrad concept one extracts a unique type D solution in vacuum with $\Lambda$. In fact, this type D solution is a limitation in vacuum with $\Lambda$ of Debever's solution \cite{debever1984exhaustive} and Papakostas' solution \cite{papakostas1998generalization}, wherein both solutions were obtained based on the aforementioned concept.  

Although, applying the methodology of Chapter 3 we result in two conformally flat solutions \footnote{The proof of this statement is not given here, although it is easy for the reader to verify this using the IC of the \textit{Paradigm} (relations (11) in \cite{papakostas1998generalization}).}, wherein the implication of the null rotation just annihilates the parameter $p$. Basically this outcome shows that we obtain more general Petrov types solutions (type D) abolishing the arbitrariness of the tetrad. Subsequently, a question arises, which is the most general transformation that we must apply in order to acquire the most general Petrov type solution?    

Moving forward, another important matter was brought to surface by this study, is that the canonical forms appear to be more fruitful instead of the Killing tensor of our \textit{Paradigm} satisfying the conjecture of Hauser-Malhiot \cite{Hauser1976}. This is evident by the vast variety of the different Petrov types were obtained assuming the $K^2, K^3$ Killing forms and $K^4$ form by extension \footnote{Once more, we denote that $K^4$ form is not a typical canonical form but it is originated by the reduction of $K^1, K^2, K^3$ where $\lambda_0=0$.}. Also, in reference to the work of Van den Bergh \cite{van2017algebraically} our type D solutions, one with a shear-free and non-geodesic null congruence (Theorem \ref{theorem 4}) and the other with a shearing and geodesic null congruence (Theorem \ref{theorem 5}) might be new type D solutions. In fact, they could be possible reductions (in vacuum with $\Lambda$) of the undiscovered solutions with PND of Einstein-Maxwell tensor aligned (or non-aligned) with the PND of Weyl tensor. 

Finally, we hope that a similar study of the canonical forms in electro-vacuum with $\Lambda$ will be able to address most of the questions posed. \textbf{In this future work, we also aim to investigate whether the generality of the canonical forms of the Killing tensor leads to more general canonical forms of Weyl tensor}.

\bibliography{sn-bibliography}


\begin{thebibliography}{36}
\ifx \bisbn   \undefined \def \bisbn  #1{ISBN #1}\fi
\ifx \binits  \undefined \def \binits#1{#1}\fi
\ifx \bauthor  \undefined \def \bauthor#1{#1}\fi
\ifx \batitle  \undefined \def \batitle#1{#1}\fi
\ifx \bjtitle  \undefined \def \bjtitle#1{#1}\fi
\ifx \bvolume  \undefined \def \bvolume#1{\textbf{#1}}\fi
\ifx \byear  \undefined \def \byear#1{#1}\fi
\ifx \bissue  \undefined \def \bissue#1{#1}\fi
\ifx \bfpage  \undefined \def \bfpage#1{#1}\fi
\ifx \blpage  \undefined \def \blpage #1{#1}\fi
\ifx \burl  \undefined \def \burl#1{\textsf{#1}}\fi
\ifx \doiurl  \undefined \def \doiurl#1{\url{https://doi.org/#1}}\fi
\ifx \betal  \undefined \def \betal{\textit{et al.}}\fi
\ifx \binstitute  \undefined \def \binstitute#1{#1}\fi
\ifx \binstitutionaled  \undefined \def \binstitutionaled#1{#1}\fi
\ifx \bctitle  \undefined \def \bctitle#1{#1}\fi
\ifx \beditor  \undefined \def \beditor#1{#1}\fi
\ifx \bpublisher  \undefined \def \bpublisher#1{#1}\fi
\ifx \bbtitle  \undefined \def \bbtitle#1{#1}\fi
\ifx \bedition  \undefined \def \bedition#1{#1}\fi
\ifx \bseriesno  \undefined \def \bseriesno#1{#1}\fi
\ifx \blocation  \undefined \def \blocation#1{#1}\fi
\ifx \bsertitle  \undefined \def \bsertitle#1{#1}\fi
\ifx \bsnm \undefined \def \bsnm#1{#1}\fi
\ifx \bsuffix \undefined \def \bsuffix#1{#1}\fi
\ifx \bparticle \undefined \def \bparticle#1{#1}\fi
\ifx \barticle \undefined \def \barticle#1{#1}\fi
\bibcommenthead
\ifx \bconfdate \undefined \def \bconfdate #1{#1}\fi
\ifx \botherref \undefined \def \botherref #1{#1}\fi
\ifx \url \undefined \def \url#1{\textsf{#1}}\fi
\ifx \bchapter \undefined \def \bchapter#1{#1}\fi
\ifx \bbook \undefined \def \bbook#1{#1}\fi
\ifx \bcomment \undefined \def \bcomment#1{#1}\fi
\ifx \oauthor \undefined \def \oauthor#1{#1}\fi
\ifx \citeauthoryear \undefined \def \citeauthoryear#1{#1}\fi
\ifx \endbibitem  \undefined \def \endbibitem {}\fi
\ifx \bconflocation  \undefined \def \bconflocation#1{#1}\fi
\ifx \arxivurl  \undefined \def \arxivurl#1{\textsf{#1}}\fi
\csname PreBibitemsHook\endcsname

\bibitem[\protect\citeauthoryear{Churchill}{1932}]{churchill1932canonical}
\begin{barticle}
\bauthor{\bsnm{Churchill}, \binits{R.V.}}:
\batitle{Canonical forms for symmetric linear vector functions in
  {pseudo-Euclidean space}}.
\bjtitle{Trans. Amer. Math. Soc.}
\bvolume{34}(\bissue{4}),
\bfpage{784}--\blpage{794}
(\byear{1932})
\end{barticle}
\endbibitem

\bibitem[\protect\citeauthoryear{Rainich}{1925}]{rainich1925electrodynamics}
\begin{barticle}
\bauthor{\bsnm{Rainich}, \binits{G.Y.}}:
\batitle{Electrodynamics in the general relativity theory}.
\bjtitle{Transactions of the American Mathematical Society}
\bvolume{27}(\bissue{1}),
\bfpage{106}--\blpage{136}
(\byear{1925})
\end{barticle}
\endbibitem

\bibitem[\protect\citeauthoryear{Newman and Penrose}{1962}]{Newman1962}
\begin{barticle}
\bauthor{\bsnm{Newman}, \binits{E.}},
\bauthor{\bsnm{Penrose}, \binits{R.}}:
\batitle{An approach to gravitational radiation by a method of spin
  coefficients}.
\bjtitle{J. Math. Phys.}
\bvolume{3}(\bissue{3}),
\bfpage{566}--\blpage{578}
(\byear{1962})
\end{barticle}
\endbibitem

\bibitem[\protect\citeauthoryear{Petrov}{2000}]{petrov2000classification}
\begin{barticle}
\bauthor{\bsnm{Petrov}, \binits{A.Z.}}:
\batitle{The classification of spaces defining gravitational fields}.
\bjtitle{Gen. Rel. Grav.}
\bvolume{32}(\bissue{8}),
\bfpage{1665}--\blpage{1685}
(\byear{2000})
\end{barticle}
\endbibitem

\bibitem[\protect\citeauthoryear{{Van den Bergh}}{2017}]{van2017algebraically}
\begin{barticle}
\bauthor{\bsnm{{Van den Bergh}}, \binits{N.}}:
\batitle{Algebraically special {Einstein-Maxwell} fields}.
\bjtitle{Gen. Rel. Grav.}
\bvolume{49}(\bissue{1}),
\bfpage{9}
(\byear{2017})
\end{barticle}
\endbibitem

\bibitem[\protect\citeauthoryear{Hauser and Malhiot}{1976}]{Hauser1976}
\begin{barticle}
\bauthor{\bsnm{Hauser}, \binits{I.}},
\bauthor{\bsnm{Malhiot}, \binits{R.J.}}:
\batitle{On space-time {Killing tensors with a Segr\'e} characteristic
  [(11),(11)]}.
\bjtitle{J. Math. Phys.}
\bvolume{17}(\bissue{7}),
\bfpage{1306}--\blpage{1312}
(\byear{1976})
\end{barticle}
\endbibitem

\bibitem[\protect\citeauthoryear{Hauser and Malhiot}{1978}]{hauser1978forms}
\begin{barticle}
\bauthor{\bsnm{Hauser}, \binits{I.}},
\bauthor{\bsnm{Malhiot}, \binits{R.J.}}:
\batitle{Forms of all spacetime metrics which admit [(11)(11)] {Killing}
  tensors with nonconstant eigenvalues}.
\bjtitle{J. Math. Phys.}
\bvolume{19}(\bissue{1}),
\bfpage{187}--\blpage{194}
(\byear{1978})
\end{barticle}
\endbibitem

\bibitem[\protect\citeauthoryear{Papakostas}{1998}]{papakostas1998generalization}
\begin{barticle}
\bauthor{\bsnm{Papakostas}, \binits{T.}}:
\batitle{{A Generalization of the Wahlquist Solution}}.
\bjtitle{Int. J. Mod. Phys. D}
\bvolume{7}(\bissue{06}),
\bfpage{927}--\blpage{941}
(\byear{1998})
\end{barticle}
\endbibitem

\bibitem[\protect\citeauthoryear{Carter}{1968}]{carter1968hamilton}
\begin{barticle}
\bauthor{\bsnm{Carter}, \binits{B.}}:
\batitle{{Hamilton-Jacobi and Schrodinger separable solutions of Einstein’s}
  equations}.
\bjtitle{Comm. Math. Phys.}
\bvolume{10},
\bfpage{280}--\blpage{310}
(\byear{1968})
\end{barticle}
\endbibitem

\bibitem[\protect\citeauthoryear{Burns and Matveev}{2021}]{burns2021open}
\begin{barticle}
\bauthor{\bsnm{Burns}, \binits{K.}},
\bauthor{\bsnm{Matveev}, \binits{V.S.}}:
\batitle{Open problems and questions about geodesics}.
\bjtitle{Ergodic Theory and Dynamical Systems}
\bvolume{41}(\bissue{3}),
\bfpage{641}--\blpage{684}
(\byear{2021})
\end{barticle}
\endbibitem

\bibitem[\protect\citeauthoryear{Eisenhart}{1934}]{eisenhart1934separable}
\begin{botherref}
\oauthor{\bsnm{Eisenhart}, \binits{L.P.}}:
{Separable systems of St\"ackel}.
Annals of Mathematics,
284--305
(1934)
\end{botherref}
\endbibitem

\bibitem[\protect\citeauthoryear{Kalnins and Miller}{1980}]{kalnins1980killing}
\begin{barticle}
\bauthor{\bsnm{Kalnins}, \binits{E.G.}},
\bauthor{\bsnm{Miller}, \binits{W.} \bsuffix{Jr}}:
\batitle{{Killing tensors and variable separation for Hamilton-Jacobi and
  Helmholtz equations}}.
\bjtitle{SIAM Journal on Mathematical Analysis}
\bvolume{11}(\bissue{6}),
\bfpage{1011}--\blpage{1026}
(\byear{1980})
\end{barticle}
\endbibitem

\bibitem[\protect\citeauthoryear{Kalnins and Miller}{1981}]{kalnins1981killing}
\begin{barticle}
\bauthor{\bsnm{Kalnins}, \binits{E.G.}},
\bauthor{\bsnm{Miller}, \binits{W.} \bsuffix{Jr}}:
\batitle{Killing tensors and nonorthogonal variable separation for
  {Hamilton-Jacobi equations}}.
\bjtitle{SIAM Journal on Mathematical Analysis}
\bvolume{12}(\bissue{4}),
\bfpage{617}--\blpage{629}
(\byear{1981})
\end{barticle}
\endbibitem

\bibitem[\protect\citeauthoryear{Kalnins and
  Miller}{1983}]{kalnins1983conformal}
\begin{barticle}
\bauthor{\bsnm{Kalnins}, \binits{E.G.}},
\bauthor{\bsnm{Miller}, \binits{W.} \bsuffix{Jr}}:
\batitle{Conformal killing tensors and variable separation for
  {Hamilton-Jacobi} equations}.
\bjtitle{SIAM Journal on Mathematical Analysis}
\bvolume{14}(\bissue{1}),
\bfpage{126}--\blpage{137}
(\byear{1983})
\end{barticle}
\endbibitem

\bibitem[\protect\citeauthoryear{Benenti}{2016}]{benenti2016separability}
\begin{barticle}
\bauthor{\bsnm{Benenti}, \binits{S.}}:
\batitle{{Separability in Riemannian manifolds}}.
\bjtitle{SIGMA. Symmetry, Integrability and Geometry: Methods and Applications}
\bvolume{12},
\bfpage{013}
(\byear{2016})
\end{barticle}
\endbibitem

\bibitem[\protect\citeauthoryear{Papakostas}{2001}]{Papakostas2001}
\begin{barticle}
\bauthor{\bsnm{Papakostas}, \binits{T.}}:
\batitle{Anisotropic fluids in the case of stationary and axisymmetric spaces
  of {General Relativity}}.
\bjtitle{Int. J. Mod. Phys. D}
\bvolume{10}(\bissue{06}),
\bfpage{869}--\blpage{879}
(\byear{2001})
\end{barticle}
\endbibitem

\bibitem[\protect\citeauthoryear{Cariglia}{2014}]{cariglia2014hidden}
\begin{barticle}
\bauthor{\bsnm{Cariglia}, \binits{M.}}:
\batitle{Hidden symmetries of dynamics in classical and quantum physics}.
\bjtitle{Rev. Mod. Phys.}
\bvolume{86}(\bissue{4}),
\bfpage{1283}
(\byear{2014})
\end{barticle}
\endbibitem

\bibitem[\protect\citeauthoryear{Papakostas}{1985}]{taxiarchis1985space}
\begin{barticle}
\bauthor{\bsnm{Papakostas}, \binits{T.}}:
\batitle{Space-times admitting {Penrose-Floyd} tensors}.
\bjtitle{Gen. Rel. Grav.}
\bvolume{17},
\bfpage{149}--\blpage{166}
(\byear{1985})
\end{barticle}
\endbibitem

\bibitem[\protect\citeauthoryear{Carter}{1968}]{Carter1968b}
\begin{barticle}
\bauthor{\bsnm{Carter}, \binits{B.}}:
\batitle{Global structure of the {Kerr} family of gravitational fields}.
\bjtitle{Phys. Rev.}
\bvolume{174}(\bissue{5}),
\bfpage{1559}
(\byear{1968})
\end{barticle}
\endbibitem

\bibitem[\protect\citeauthoryear{Cahen et~al.}{1967}]{cahen1967complex}
\begin{barticle}
\bauthor{\bsnm{Cahen}, \binits{M.}},
\bauthor{\bsnm{Debever}, \binits{R.}},
\bauthor{\bsnm{Defrise}, \binits{L.}}:
\batitle{{A Complex Vectorial Formalism in General Relativity}}.
\bjtitle{Journal of Mathematics and Mechanics}
\bvolume{16}(\bissue{7}),
\bfpage{761}--\blpage{785}
(\byear{1967})
\end{barticle}
\endbibitem

\bibitem[\protect\citeauthoryear{Newman and Penrose}{1962}]{newman1962approach}
\begin{barticle}
\bauthor{\bsnm{Newman}, \binits{E.}},
\bauthor{\bsnm{Penrose}, \binits{R.}}:
\batitle{An approach to gravitational radiation by a method of spin
  coefficients}.
\bjtitle{J. Math. Phys.}
\bvolume{3}(\bissue{3}),
\bfpage{566}--\blpage{578}
(\byear{1962})
\end{barticle}
\endbibitem

\bibitem[\protect\citeauthoryear{Debever}{1964}]{debeverriemann}
\begin{barticle}
\bauthor{\bsnm{Debever}, \binits{R.}}:
\batitle{Le rayonnement gravitationnel}.
\bjtitle{Cahiers de Physique}
\bvolume{8},
\bfpage{303}--\blpage{349}
(\byear{1964})
\end{barticle}
\endbibitem

\bibitem[\protect\citeauthoryear{Stephani et~al.}{2009}]{stephani2009exact}
\begin{bbook}
\bauthor{\bsnm{Stephani}, \binits{H.}},
\bauthor{\bsnm{Kramer}, \binits{D.}},
\bauthor{\bsnm{MacCallum}, \binits{M.}},
\bauthor{\bsnm{Hoenselaers}, \binits{C.}},
\bauthor{\bsnm{Herlt}, \binits{E.}}:
\bbtitle{Exact Solutions of {Einstein's} Field Equations}.
\bpublisher{{Cambridge University Press}},
\blocation{NY}
(\byear{2009})
\end{bbook}
\endbibitem

\bibitem[\protect\citeauthoryear{Stewart}{1993}]{stewart1993advanced}
\begin{bbook}
\bauthor{\bsnm{Stewart}, \binits{J.}}:
\bbtitle{Advanced General Relativity}.
\bpublisher{{Cambridge University Press}},
\blocation{NY}
(\byear{1993})
\end{bbook}
\endbibitem

\bibitem[\protect\citeauthoryear{Kruglikov and
  Matveev}{2016}]{kruglikov2016geodesic}
\begin{barticle}
\bauthor{\bsnm{Kruglikov}, \binits{B.}},
\bauthor{\bsnm{Matveev}, \binits{V.S.}}:
\batitle{The geodesic flow of a generic metric does not admit nontrivial
  integrals polynomial in momenta}.
\bjtitle{Nonlinearity}
\bvolume{29}(\bissue{6}),
\bfpage{1755}
(\byear{2016})
\end{barticle}
\endbibitem

\bibitem[\protect\citeauthoryear{Sommers}{1973}]{sommers1973killing}
\begin{barticle}
\bauthor{\bsnm{Sommers}, \binits{P.}}:
\batitle{On {Killing tensors} and constants of motion}.
\bjtitle{J. Math. Phys.}
\bvolume{14}(\bissue{6}),
\bfpage{787}--\blpage{790}
(\byear{1973})
\end{barticle}
\endbibitem

\bibitem[\protect\citeauthoryear{Sadeghian}{2022}]{sadeghian2022killing}
\begin{barticle}
\bauthor{\bsnm{Sadeghian}, \binits{S.}}:
\batitle{Killing tensors of a generalized {Lense-Thirring} spacetime}.
\bjtitle{Phys. Rev. D.}
\bvolume{106}(\bissue{10}),
\bfpage{104028}
(\byear{2022})
\end{barticle}
\endbibitem

\bibitem[\protect\citeauthoryear{Garfinkle and
  Glass}{2010}]{garfinkle2010killing}
\begin{barticle}
\bauthor{\bsnm{Garfinkle}, \binits{D.}},
\bauthor{\bsnm{Glass}, \binits{E.N.}}:
\batitle{Killing tensors and symmetries}.
\bjtitle{Class. Quantum. Grav.}
\bvolume{27}(\bissue{9}),
\bfpage{095004}
(\byear{2010})
\end{barticle}
\endbibitem

\bibitem[\protect\citeauthoryear{Frolov et~al.}{2017}]{frolov2017black}
\begin{barticle}
\bauthor{\bsnm{Frolov}, \binits{V.P.}},
\bauthor{\bsnm{Krtou{\v{s}}}, \binits{P.}},
\bauthor{\bsnm{Kubiz{\v{n}}{\'a}k}, \binits{D.}}:
\batitle{Black holes, hidden symmetries, and complete integrability}.
\bjtitle{Liv. Rev. Rel.}
\bvolume{20},
\bfpage{1}--\blpage{221}
(\byear{2017})
\end{barticle}
\endbibitem

\bibitem[\protect\citeauthoryear{Krtou{\v{s}}
  et~al.}{2007}]{krtouvs2007killing}
\begin{barticle}
\bauthor{\bsnm{Krtou{\v{s}}}, \binits{P.}},
\bauthor{\bsnm{Kubizn{\'a}k}, \binits{D.}},
\bauthor{\bsnm{Page}, \binits{D.N.}},
\bauthor{\bsnm{Frolov}, \binits{V.P.}}:
\batitle{{Killing-Yano tensors, rank-2 Killing tensors, and conserved
  quantities in higher dimensions}}.
\bjtitle{JHEP}
\bvolume{2007}(\bissue{02}),
\bfpage{004}
(\byear{2007})
\end{barticle}
\endbibitem

\bibitem[\protect\citeauthoryear{Benenti and
  Francaviglia}{1979}]{benenti1979remarks}
\begin{barticle}
\bauthor{\bsnm{Benenti}, \binits{S.}},
\bauthor{\bsnm{Francaviglia}, \binits{M.}}:
\batitle{Remarks on certain separability structures and their applications to
  {General Relativity}}.
\bjtitle{Gen. Rel. Grav.}
\bvolume{10},
\bfpage{79}--\blpage{92}
(\byear{1979})
\end{barticle}
\endbibitem

\bibitem[\protect\citeauthoryear{Landau and
  Lifschitz}{1975}]{landau2013classical}
\begin{bbook}
\bauthor{\bsnm{Landau}, \binits{L.D.}},
\bauthor{\bsnm{Lifschitz}, \binits{E.M.}}:
\bbtitle{{The Classical Theory of Fields}}
vol. \bseriesno{II}.
\bpublisher{Pergamon},
\blocation{Oxford}
(\byear{1975})
\end{bbook}
\endbibitem

\bibitem[\protect\citeauthoryear{Debever et~al.}{1984}]{debever1984nouvelle}
\begin{bchapter}
\bauthor{\bsnm{Debever}, \binits{R.}},
\bauthor{\bsnm{Kamran}, \binits{N.}},
\bauthor{\bsnm{McLenaghan}, \binits{R.G.}}:
\bctitle{{Sur une nouvelle expression de la solution g{\'e}n{\'e}rale des
  {\'e}quations d'Einstein avec champ de {Maxwell} non singulier, align{\'e},
  sans source et avec constante cosmologique, en type D}}.
In: \bbtitle{Annales de l'IHP Physique Th{\'e}orique},
vol. \bseriesno{41},
pp. \bfpage{191}--\blpage{206}
(\byear{1984})
\end{bchapter}
\endbibitem

\bibitem[\protect\citeauthoryear{Strang}{2022}]{strang2022introduction}
\begin{bbook}
\bauthor{\bsnm{Strang}, \binits{G.}}:
\bbtitle{Introduction to Linear Algebra}.
\bpublisher{SIAM},
\blocation{Wellesley}
(\byear{2022})
\end{bbook}
\endbibitem

\bibitem[\protect\citeauthoryear{Debever and
  McLenaghan}{1981}]{debever1981orthogonal}
\begin{barticle}
\bauthor{\bsnm{Debever}, \binits{R.}},
\bauthor{\bsnm{McLenaghan}, \binits{R.G.}}:
\batitle{{Orthogonal transitivity, invertibility, and null geodesic
  separability in type D electrovac solutions of Einstein’s field equations
  with cosmological constant}}.
\bjtitle{J. Math. Phys.}
\bvolume{22}(\bissue{8}),
\bfpage{1711}--\blpage{1726}
(\byear{1981})
\end{barticle}
\endbibitem

\bibitem[\protect\citeauthoryear{Debever et~al.}{1984}]{debever1984exhaustive}
\begin{barticle}
\bauthor{\bsnm{Debever}, \binits{R.}},
\bauthor{\bsnm{Kamran}, \binits{N.}},
\bauthor{\bsnm{McLenaghan}, \binits{R.G.}}:
\batitle{Exhaustive integration and a single expression for the general
  solution of the type {D} vacuum and electrovac field equations with
  cosmological constant for a nonsingular aligned maxwell field}.
\bjtitle{J. Math. Phys.}
\bvolume{25}(\bissue{6}),
\bfpage{1955}--\blpage{1972}
(\byear{1984})
\end{barticle}
\endbibitem

\end{thebibliography}

\end{document}